\documentclass[12pt]{article}
\usepackage{amsmath,amsthm,amsfonts,amssymb,amsopn,amscd}     

\DeclareMathOperator{\Tr}{Tr}

\def\qed{{\unskip\nobreak\hfil\penalty50
\hskip2em\hbox{}\nobreak\hfil$\square$
\parfillskip=0pt \finalhyphendemerits=0\par}\medskip}
\def\proof{\trivlist \item[\hskip \labelsep{\bf Proof.\ }]}
\def\endproof{\null\hfill\qed\endtrivlist\noindent}

\def\aci{\underset{t\to -i}{\rm anal.cont.\,}}

\def\Ad{{\hbox{\rm Ad}}}

\def\dim{{\hbox{dim}}}

\def\o{{\omega}}

\def\a{\alpha}
\def\b{\beta}

\def\e{\varepsilon}

\def\Ga{\Gamma}
\def\k{\kappa}
\def\l{\lambda}

\def\r{\rho}
\def\phi{\varphi}

\def\Om{\Omega}

\newtheorem{theorem}{Theorem}
\newtheorem{lemma}[theorem]{Lemma}
\newtheorem{corollary}[theorem]{Corollary}

\newtheorem{proposition}[theorem]{Proposition}

\def\setminus{\smallsetminus}
\def\A{{\cal A}}
\def\B{{\cal B}}

\def\E{{\cal E}}

\def\I{{\cal I}}
\def\H{{\cal H}}

\def\f{{\varphi}}
\def\s{{\sigma}}

\def\emptyset{\varnothing}
\def\res{{\upharpoonright}}

\def\Mob{{\rm\textsf{M\"ob}}}

\def\emptyset{\varnothing}
\def\setminus{\smallsetminus}
\def\Vect{{\mathrm {Vect}}}
\def\Diff{{\mathrm {Diff}}}
\def\Mob{{\rm\textsf{M\"ob}}}

\title{{\bf
Noncommutative Spectral Invariants \\
and Black Hole Entropy}}
\medskip

\author{
{\sc Yasuyuki Kawahigashi}\footnote{Supported in part by JSPS.}\\
Department of Mathematical Sciences\\
University of Tokyo, Komaba, Tokyo, 153-8914, Japan\\
e-mail: {\tt yasuyuki@ms.u-tokyo.ac.jp}\\
\vphantom{X}\\
{\sc Roberto Longo}\footnote{Supported in part by GNAMPA and MIUR.}\\
Dipartimento di Matematica, 
Universit\`a di Roma ``Tor Vergata''\\
Via della Ricerca Scientifica, 1, I-00133 Roma, Italy\\
e-mail: {\tt longo@mat.uniroma2.it}}
\begin{document}
\date{}
\maketitle
\bigskip

\begin{abstract}
We consider an intrinsic entropy associated with a local conformal net
$\A$ by the coefficients in the expansion of the logarithm of the trace
of the ``heat kernel'' semigroup.  In analogy with Weyl theorem on the
asymptotic density distribution of the Laplacian eigenvalues, passing
to a quantum system with infinitely many degrees of freedom, we
regard these coefficients as noncommutative geometric invariants.
Under a natural modularity assumption, the leading term of the
entropy (noncommutative area) is proportional to the central charge
$c$, the first order correction (noncommutative Euler characteristic)
is proportional to $\log\mu_{\A}$, where $\mu_{\A}$ is the global
index of $\A$, and the second spectral invariant is again proportional
to $c$.

We give a further general method to define a mean entropy by considering 
conformal symmetries that preserve a discretization of $S^1$ and we 
get the same value proportional to $c$.

We then make the corresponding analysis with the proper Hamiltonian 
associated to an interval.  We find here, in complete generality, a 
proper mean entropy proportional to $\log\mu_{\A}$ with a first order 
correction defined by means of the relative entropy associated with 
canonical states.
\\*

By considering a class of black holes with an associated conformal
quantum field theory on the horizon, and relying on arguments in the
literature, we indicate a possible way to link the noncommutative area
with the Bekenstein-Hawking classical area description of entropy.

\end{abstract}
\section{Introduction}
This paper essentially deals with chiral conformal Quantum Field 
Theory, but our motivations primarily concern black hole 
thermodynamics; the basic link to this subject is through QFT on a 
curved spacetime and the idea, that has appeared from different and 
independent viewpoints in recent literature, that the restriction 
of the quantum field to the black hole horizon should give rise to a 
conformal QFT.  Combined with the well known Bekenstein interpretation 
of the area of the horizon as proportional to the black hole entropy, 
this suggests that a geometric definition of the entropy of conformal 
QFT should play a relevant r\^{o}le in black hole thermodynamics.  
To this end we shall define an intrinsic entropy associated to a conformal 
QFT, with a noncommutative geometrical point of view.
We will regard a local conformal net as a noncommutative manifold or, 
more precisely, a QFT manifold (i.e.  a noncommutative manifold with 
infinitely many degrees of freedom) and shall be guided in our 
analysis by the classical equivalent, most importantly from Weyl's 
asymptotic for the trace of the heat kernel.  One could say that in our framework 
back reaction effects of the quantum fields on the classical spacetime 
are negligible, but do affect the geometry of the associated 
noncommutative manifold.

Our paper is organized as follows:
\begin{itemize}
\item Here below we recall a number of ideas about black hole physics 
that have motivated our work, yet we refer to the literature (see e.g.  
\cite{Wa}) for basics facts on black hole thermodynamics as Hawking 
effect, generalized second low, etc..

\item We then recall Weyl's theorem that motivates our ``log-ellipticity'' 
assumption on the conformal Hamiltonian, i.e.  on the 
asymptotic of logarithmic of the characters (elementary motivations are 
contained in Appendix \ref{App.Trace}).  This assumption holds in all 
computed cases. We shall show that it holds for all modular local 
conformal nets, namely nets with the usual rational behavior (see Sect.  
\ref{si}) and it turns out to hold in particular in all models 
with central charge less than one, that are classified in \cite{KL,KL2}. 
Indeed one has the asymptotic formula for a modular net $\A$
\begin{equation}\label{spinv}
\log\Tr(e^{-2\pi tL_{0,\r}})
\sim \frac{\pi c}{12}\frac1t  + \log\frac{d(\r)}{\sqrt{\mu_{\A}}} 
-\frac{\pi c}{12} t\ ,\quad \text{as}\ t\to 0^+
\end{equation}
where $c$ is the central charge, $L_{0,\r}$ and $d(\r)$ are the
conformal Hamiltonian and the DHR dimension of the representation
$\r$, and $\mu_\A$ is equal to the global index $\sum_{i}d(\r_i)^2$,
the sum of the indices of all DHR charges \cite{KLM,LX} (see Sect.
\ref{si}).

\item Our basic object is a local conformal net $\A$ of von Neumann 
algebras, namely the family of local operator algebras maximally 
generated by smeared fields (basic notions can be found in Appendix 
\ref{nets}); this is our noncommutative manifold and we use 
(temporarily) the log-ellipticity/modularity assumption for our 
analysis.  In analogy with Weyl's theorem we define the noncommutative 
geometric spectral invariants $\{a_i\}$ of a conformal net (the 
coefficients in the above asymptotic (\ref{spinv},\ref{asymp})), in 
particular the noncommutative area and the noncommutative Euler 
characteristic.  Indeed as we are in the QFT setting (thus with 
infinitely many degrees of freedom) $\log\Tr(e^{-2\pi tL_0})$, rather 
than $\Tr(e^{-2\pi tL_0})$, provides the asymptotic of the 
corresponding finite-dimensional system, see Appendix \ref{App.Trace}.

For the Physics viewpoint, $\log\Tr(e^{-2\pi tL_0})$ 
counts logarithmically the number of possible states and so determines 
the microscopic entropy $S_\A$ of the system, therefore we put
\[
S_\A \equiv a_0 \ .
\]
The following table summarizes the value and the meaning of the 
spectral invariants (up to proportionality constants):
\[
\text{{\footnotesize\begin{tabular}{|c|c|l|l|}
{\it Invariant}&{\it Value} & {\it Geometry}& {\it 
Physics}\\
\hline
$a_0$   & $\pi c/12$   & Noncommutative area & Entropy\\
$a_1$   & $-\tfrac12\log\mu_\A$   & Noncommmutative Euler 
characteristic  & $1^{\rm st}$ order entropy\\
$a_2$   & $-\pi c/12$ & $2^{\rm nd}$ spectral invariant & $2^{\rm nd}$ order entropy\\
\hline
\end{tabular}}} 
\]
Note that $a_2 = -a_0$, that is a consequence of the modular symmetry.

The analog of the Kac-Wakimoto formula \cite{L2}, and more generally
the quantum index formula in \cite{L4}, can now be read as an
expression that the incremental free energy (adding/removing DHR charges
\cite{DHR}) is proportional to the increment of the noncommutative
Euler characteristic (Sect.  \ref{dF}).

\item We shall show that, for a conformal net on the two-dimensional 
Minkowski spacetime, an expansion analog to \eqref{spinv} holds, where 
$a_0$ duplicates.  At this point we look for a direct connections with 
black hole thermodynamics.  In the paper \cite{Ca} (following 
\cite{SV}) on black holes one finds computations that fit well with 
our results.  There $c/12 = A/8\pi$ so one immediately gets that $S_\A$ 
has the Bekenstein behavior
\[
S_\A = A/4 \ ,
\]
where $A$ is the classical area of the black hole horizon.

\item We then provide a general analysis where we do not any longer use the
modularity assumption.  We first recall how the $n$-cover $\Diff^{(n)}(S^1)$ of
$\Diff(S^1)$ acts on $S^1$, see \cite{LX}.  The generator of the
corresponding rotation one-parameter group is viewed as a conformal
Hamiltonian associated with a discretization of $S^1$, namely to a
partition of $S^1$ in $n$ intervals, where $n$ is then supposed to
tend to infinity.

If one subtracts from the corresponding entropy (logarithm of
partition function) the naive entropy associated with $1/n$ times the
original conformal Hamiltonian, the resulting entropy should take into
account the noncommutative geometrical complexity.  We thus give in
this way a general definition of mean free energy and it turns out
immediately that 
\[
F_{\rm mean}= \pi c/12\ ,
\]
that agrees with the above found value for the entropy $a_0$, hence 
again  $F_{\rm mean}= A/4$ in the above setting.

\item At this point we get in the second part of the paper, where we
study the ``local'' version of the above structure, namely we 
consider the operator algebra associated with a given interval and the 
associated proper dynamics with a one-parameter group of special 
conformal transformations.  We consider the
generators of this ``dilatation'' group in $\Diff(S^1)$ and in
$\Diff^{(n)}(S^1)$ as Hamiltonians and we attempt to compute the
associated noncommutative spectral invariants.  Only conformal
symmetries and the split property play a r\^{o}le here and results are
very general.

We then extend to the general model independent setting a formula by
Schroer and Wiesbrock \cite{SW} for the Tomita-Takesaki modular group
of the von Neumann algebra associated with $n$ separated intervals; in
other words we prove the KMS thermal equilibrium property, for above
proper dynamics associated with the discretization of $S^1$, with
respect to a canonical state, in any representation.  This is one of
our main tools for the sequel.

\item With this proper Hamiltonian, in analogy with the previous
analysis, we define the partition function $Z_n$ associated with this
discretization of $S^1$ with $n$-intervals and then the $\mu$-free
energy
$F_{\rm mean, \mu}$
as the $\lim_{n\to\infty}-\b^{-1}\log
Z_n(\b)/n$ at inverse equilibrium temperature $\b$ (Hawking
temperature).  It turns out that, in any irreducible representation,
\[
F_{\rm mean,\mu}=\frac12\log\mu_{\A}\ ,
\]
where $\mu_{\A}$ is the
$\mu$-index of the net, namely the Jones index of the 2-interval
inclusion of von Neumann algebras in the vacuum sector
\cite{KLM} (Sect.  \ref{si}).

Pursuing the above analogy we interpret the first noncommutative 
local spectral invariants. It turns out that the $0^{\rm th}$ invariant 
$a_{0,\mu}$, equal by definition to the proper noncommutative 
area, is proportional to the mean entropy. The first spectral 
invariant $a_{1,\mu}$, equal by definition to the proper noncommutative 
Euler characteristic, turns out to be proportional to the proper mean 
entropy $S_{\rm mean,\mu}$. (Locally the $\mu$-index seems to play 
the r\^{o}le of the central charge globally, but we have no definite 
interpretation of this fact.)

\item Our mathematical methods concern Jones' index \cite{J}, as extended by 
Kosaki \cite{Ko}, and Connes-Haagerup noncommutative measure theory, 
see \cite{T}. We have 
put our mathematical results in Appendix \ref{math}, in order not to 
interrupt the main theme of the paper. A quick introduction to 
Operator Algebras and Conformal Field Theory can be found in 
\cite{KLX}.
\end{itemize}

\section{On black hole entropy}
\label{BKE}
We now recall a few motivational items concerning black hole physics.

\smallskip\noindent
{\it The holographic principle} \cite{tH,Su}. The celebrated 
Bekenstein formula \cite{Be1} for the entropy of a black hole is
\[
S=\a A
\]
where $A$ is the area of the black hole horizon and $\a$ is a constant. This 
formula was initially 
motivated by consistency arguments and the area theorem. One of the 
most surprising fact is that it sets the entropy to be 
proportional to the area, rather than to the volume, as an intuitive 
picture of the entropy as logarithmic counting of the number 
of possible states would suggest.

This dimensional reduction has more recently led to the formulation of
the holographic principle according to which, in a theory combining
quantum theory and gravity, the degrees of freedom of a three
dimensional world can be stored in a two dimensional projection.  One
of the argument is that ``one can't hide behind a black hole'': if
black hole projects itself on a screen, due to gravity a second black
hole can't eclipse its image on the screen \cite{Su}.

\smallskip

\noindent
{\it Hawking temperature. Fixing the proportionality constant}. Let's 
recall how the proportionality constant can be fixed as $\a=1/4$ by 
considering quantum effects (cf. \cite{W}). As 
shown by Hawking, a black hole emits a thermal radiation with 
inverse temperature
\[
\b =\frac{2\pi}{\k}\ ,
\]
where $\k$ is the surface gravity. Let's consider 
the Schwarzschild spacetime with radius $R$, thus describing a black 
hole of mass $M=2R$. In this case $\k=\frac{1}{4M}$, thus 
$\b=8\pi M$. As
\[
S= \a A=\a 4\pi R^2 =\a 16\pi M^2
\]
we have
\[
\textrm{d}S=\a 32\pi M\textrm{d}M .
\]
On the other hand by the generalized second principle of 
thermodynamics
\[
\textrm{d}S=\b\textrm{d}H=\b\textrm{d}M\ ,
\]
where $H=M$ is the energy, so $\b=8\pi M=\a 32\pi M$ yielding
\[\a  =1/4 .
\]
{\it Limit of information. Discretization of the horizon} \cite{Be}.
Consider the horizon to be made by cells of area $\sim\ell^2$, where $\ell$ is the 
Planck length. Thus
\[
A= n\ell^2 .
\]
Now say that each cell has $k$ degrees of freedom: in the simplest example 
each cell is occupied by a particle with spin up/spin down and so 
$k=2$. The total number of degrees of freedom is then
\begin{equation}
	\textrm{Degrees of freedom} = k^n ;
\end{equation}
thus
\begin{equation}\label{discr.entr}
	\textrm{Entropy} = Cn\log k=C\frac{A}{\ell^2}\log k
\end{equation}
where $C$ is a constant, 
namely the entropy is proportional to the area $A$ of the black hole.

It follows that the increment of entropy by adding a particle to the 
black hole 
\begin{equation}\label{KW}
\textrm{d}S= C\log k
\end{equation}
is proportional to the logarithm of an integer.
More generally if there are distinct particles $p_1,p_2,\dots p_s$ and 
$p_i$ has $k_i$ degrees of freedom we have
\begin{equation}
	\textrm{Degrees of freedom} = k_1^{n_1}k_2^{n_2}\cdots k_s^{n_s} ,
\end{equation}
where $n=n_1 + n_2 + \dots n_s$, so
\begin{equation}
	\textrm{Entropy} = C\log k_1^{n_1}k_2^{n_2}\cdots k_s^{n_s} =
	C \sum_i n_i\log k_i \ .
\end{equation}

\noindent
{\it The conformal horizon of a black hole.} The horizon of a black 
hole is the boundary of the no escape region of the spacetime where 
signals can enter, but cannot get out. There is no particular physical
phenomena occurring on the horizon, an observer can cross it without 
feeling anything, yet it is a codimension one 
submanifold where certain parameters (coordinates) pick critical 
values. For this reason it is thus natural to expect the horizon to exhibit further 
symmetries acquainted at these critical values.

This point, related to the above mentioned holographic principle, is 
well expressed in the holography that holds in the anti-de Sitter 
spacetime \cite{Ma}. Here the algebraic approach gives a natural 
``coordinate free'' description \cite{Re}. More recently a general algebraic 
holography has been realized in the two-dimensional de Sitter 
spacetime by means of local conformal (pseudo)-nets of von Neumann 
algebras on $S^1$ \cite{GL5}. 

There is an apparent conflict between the discretization of  
the boundary and conformal invariance: our point of view is that the 
conformal symmetries that respect the discretization are the 
physically relevant ones. One should think of conformal 
QFT on the boundary as a noncommutative manifold, and we shall soon be 
back on this point. The corresponding structure will be 
explained later on.
\smallskip

\noindent
{\it Entropy from conformal boundary.} 
This point of view has emerged in recent years in 
different works as in \cite{SV,Ca,As} where conformal symmetries on the 
horizon are used to compute black hole entropy. 

For example, in the reference  \cite{Ca} by Carlip the black hole is 
described, in particular, by a spacetime with a (local) Killing 
horizon; a natural set of boundary conditions leads to a 
representation of the Virasoro algebra with central charge $c$ and it 
is argued that, in normalized units,
\begin{equation}\label{Carlip}
	\frac{c}{12} = \frac{A}{8\pi}\ ,
\end{equation}
where $A$ is the area of a cross section of the horizon (the black 
hole area). 
One then use a heuristic formula derived with certain assumption by Cardy
\begin{equation}\label{cardy}
\r(\l)\sim \exp\left(2\pi\sqrt{\tfrac16 c(\l - \tfrac{1}{24} c)}\right)
\quad \text{as} \ \l\to +\infty
\end{equation}
on the number of states $\r(\l)$ corresponding to the
eigenvalue $\l$ of the (two-dimensional) 
conformal Hamiltonian. One computes the boundary term  of the 
energy (that turns out to be equal to
$=A/8\pi$), inserts this and the value of $c$ in eq. (\ref{cardy}) 
and gets the expected Bekenstein behavior
\[
\log \r \sim\frac{A}{4}\ .
\]
\smallskip

\noindent {\it Operator algebras and conformal boundary.  Quantum
index theorem}.  Recall now the work in \cite{GLRV,L3} in the context
of black holes described by a curved spacetime with a bifurcate
Killing horizon.  $\A$ is a conformal net arising on the horizon.  By
applying a general theorem by Wiesbrock, $\A$ is a M\"{o}bius
covariant net (cf.  \cite{SuV,L3,L4,S}); moreover $\A$ is expected to be
diffeomorphism covariant and the diffeomorphism symmetry uniquely 
determined (see \cite{We}); for example this is the case
when the quantum field is free, as $\A$ is then isomorphic to the net
associated with the $U(1)$-current algebra, see \cite{GLRV} (this fact
has been noticed again in \cite{MP}).  We thus assume $\A$ to
be diffeomorphism covariant.

In \cite{L2,L3,L4} one obtained a general, model independent 
formula for a black hole with a bifurcate Killing horizon (assuming  
the KMS property for geodesic observers):
\begin{equation}\label{KW2}
{\rm d}F = 
\frac{2\pi}{\kappa}\big(\log d(\r) - \log d(\s)\big) 
\end{equation}
where ${\rm d}F$ is the incremental free energy by adding/removing DHR
charges $\r$, $\s$ localizable in bounded regions (\cite{DHR}),
$\k/2\pi$ is the Hawking temperature with $\k$ the surface gravity,
$d(\r)$ is the Doplicher-Haag-Roberts statistical dimension of $\r$,
that turns out to be equal to the square root of the Jones index of
$\r$ \cite{L5}.  Recall that, in a $n$-dimensional spacetimes, $n\geq
3$, we have $d(\r)\in\mathbb N\cup\infty$.  The above formula holds
also for finitely many charges, and we regard (\ref{discr.entr}) as a
physical description of \eqref{KW2}.  It can be read as a quantum
index theorem (or, more appropriately, ``QFT index theorem'' as it
concerns infinitely many degrees of freedom) where the quantum
Fredholm index $\log d(\r) - \log d(\s)$ is expressed in terms of
${\rm d}F$ and the geometric quantity $\kappa$.  A good illustration
of this point is provided by the topological sectors in \cite{LX}.
\section{QFT, heat kernel asymptotic and entropy}
\subsection{Weyl's theorem and ellipticity}
\label{Weylsection}
Let $M$ be a compact oriented Riemann manifold and $\Delta$ the Laplace 
operator on $L^2(M)$. The eigenvalues of $M$ can be thought as ``resonant 
frequencies'' of $M$ and capture most of the geometry of $M$ \cite{Ka}.

At the root of this analysis is the famous Weyl's theorem on the 
asymptotic density distribution of such eigenvalues. This can be 
stated as an asymptotic formula for the heat kernel, see \cite{Ro}.
One has the following asymptotic 
expansion as $t\to 0^+$:
 \begin{equation}
 	\Tr(e^{-t\Delta}) \sim \frac{1}{(4\pi t)^{n/2}}(a_0 + a_1 t + 
 	\cdots)
 	\label{Weyl}
 \end{equation} 
 and thus, by Tauberian theorems (see \cite{BGT}), the asymptotic formula as 
 $\l\to +\infty$
 \[
 N(\l)\sim \frac{{\rm vol}(M)}{(4\pi)^{n/2}\Gamma((n/2)+1)}\l^{n/2}
 \]
 for the number $N(\l)$ of eigenvalues of $\Delta$ less than $\l$, 
 where $\Gamma$ is Euler Gamma-function.
 
In (\ref{Weyl}) the spectral invariants $n$ and $a_0,a_1,\dots$ encode 
geometric information and in particular $n = {\rm dim}(M)$ and
\[
a_0 = {\rm vol}(M),\qquad a_1 = \frac{1}{6}\int_M \kappa(m)d{\rm 
vol}(m),
\]
where $\kappa$ is the scalar curvature, thus in particular if $n=2$ 
then $a_1$ is proportional to the Euler characteristic equal
 $\frac{1}{2\pi}\int_M \kappa(m)d{\rm vol}(m)$ by Gauss-Bonnet theorem.

Motivated by the Weyl asymptotic (\ref{Weyl}), having in mind a ``second quantized'' 
Hamiltonian (see Sect. \ref{App.Trace}, in particular Lemma 
\ref{logasymp}), we give the following definition to 
capture the asymptotic associated with the (here undefined) 
``one-particle Hamiltonian''.

A positive linear operator $H$ on a Hilbert space is \emph{log-elliptic} 
if there exists $n>0$ and $a_i\in\mathbb R$, $a_0\neq 0$, such that
\begin{equation}	\label{QE}
 	\log\Tr(e^{-tH}) \sim \frac{1}{t^{n/2}}(a_0 + a_1 t + 
 	\cdots)\quad {\rm as}\ t\to 0^+ \ .
  \end{equation}
Then 
\[
n=-2\lim_{t\to 0^+}\frac{\log\log\Tr(e^{-tH})}{\log t}
\]
is called the \emph{dimension} of $H$ and $a_i\equiv a_i(H)$ 
the $i^{\rm th}$ \emph{spectral invariant} of $H$.
The following is obvious.
\begin{lemma}\label{deltachi}
Let $H$, $H'$ be log-elliptic positive linear operators with dimension $n$ and 
$n'$ and spectral invariants $a_i$ and $a'_i$. If
\[
\lim_{t\to 0^+}\frac{\Tr(e^{-t H})}
{\Tr(e^{-t H'})}=\l\neq 0
\]
exists, then $n=n'$ and  $a_i=a'_i$, $i=0,1,2,\dots 
,m-1$,  $m\equiv \lfloor n/2\rfloor$; if $n/2$ is an integer then $\log\l=a_m - a'_m$. 
\end{lemma}
\proof
We have
\begin{multline}
\log\l=\lim_{t\to 0^+}\log\frac{\Tr(e^{-t H})}
{\Tr(e^{-t H'})}=\lim_{t\to 0^+}
\big(\log \Tr(e^{-t H}) -\log \Tr(e^{-t H'})\big)\\
=\lim_{t\to 0^+}\bigg(\frac{1}{t^{n/2}}(a_0 + a_1 t +a_2 t^2 +\dots)
-\frac{1}{t^{n'/2}}(a'_0 + a'_1 t +a'_2 t^2 +\dots)\bigg)
\end{multline}
which is possible only in the stated case.
\endproof 
\subsection{Spectral invariants associated with $L_0$}
\label{si}
The asymptotic of the character $\Tr(e^{-2\pi tL_0})$ as $t\to 0^+$ is
known for an irreducible representation of the Virasoro algebra 
\cite{W}, but is unknown for a general reducible representation, in 
particular for the representation associated with an arbitrary local conformal net.
Cardy has provided an argument based on modular invariance that 
implies
\[
\log\Tr(e^{-2\pi tL_0}) \sim {\rm const.}\frac{1}{t}\quad {\rm as}\ \ t\to 0^+
\]
where the constant depends on the central charge $c$ only.

Motivated by Weyl's theorem and the above expansion, we shall define a local conformal 
net $\A$ to be two-dimensional \emph{log-elliptic} (or QFT-elliptic)
if its conformal Hamiltonian 
$L_0$ is log-elliptic with dimension 2, see Section 
\ref{Weylsection}, namely
\begin{equation}\label{asymp}
\log\Tr(e^{-2\pi tL_0}) \sim\frac{1}{t}(a_0 + a_1 t + \cdots)
\quad {\rm as}\ \ t\to 0^+
\end{equation}
log-ellipticity is essentially the \emph{nuclearity 
condition} of Buchholz and Wichmann \cite{BW} (and we fix the dimension). 

We shall then regard $\A$ as a 2-dimensional noncommutative manifold,
where $L_0$ corresponds to the Laplacian and the spectral invariants
of $L_0$ are noncommutative geometric invariants for $\A$.  In
particular $a_0\equiv a_0(2\pi L_0)$ is $1/4\pi$ times the
\emph{noncommutative area} of $\A$ and $12a_1$ is the
\emph{noncommutative Euler characteristic} of $\A$.
\footnote{For simplicity we do not 
put a factor $1/4\pi$ in defining the asymptotic (\ref{asymp}).} Of 
course $a_0, a_1, \dots$ have a priori no classical geometric 
interpretation, but are defined in analogy with classical invariants.

We now explain how to obtain a more precise form of the asymptotic 
\eqref{asymp} under a general condition. Let $\A$ be a
completely rational local conformal field net on $S^1$.
For a DHR sector $\r$, we consider the specialized character
$\chi_\r(\tau)$ for complex numbers $\tau$ with ${\rm Im}\; \tau >0$
as follows:
\[
\chi_\r(\tau)=\Tr\big(e^{2\pi i\tau (L_{0,\r}-c/24)}\big).
\]
Here the operator $L_{0,\r}$ is the conformal Hamiltonian in 
the representation $\r$ and $c$ is
the central charge.  We assume that the above Trace converges, which
in particular means each eigenspace of $L_{0,\r}$ is finite
dimensional.  On one hand, it is known in many cases
that we have an action
of $SL(2,{\mathbb Z})$ on the linear span of these specialized
characters through change of variables $\tau$ as follows:
\begin{equation}\label{modularity}
\begin{split}
\chi_\r(-1/\tau)&= \sum_\nu S^\chi_{\r,\nu} \chi_\nu(\tau),\\
\chi_\r(\tau+1)&= \sum_\nu T^\chi_{\r,\nu} \chi_\nu(\tau).
\end{split}
\end{equation}
On the other hand, we have a unitary representation of the group
$SL(2,{\mathbb Z})$ on the space spanned by the sector $\r$'s
arising from the nondegenerate braiding as in
Rehren \cite{R1}, in particular we have the associated
matrices $(S_{\r,\nu})$ and
$(T_{\r,\nu})$.  It has been conjectured, e.g. 
Fr\"ohlich-Gabbiani \cite[page 625]{FG}, that these two representations
coincide, that is, we have $S^\chi=S$, $T^\chi=T$.  Note that
we always have $T^\chi=T$ by the spin-statistics theorem
\cite{GL2}, so in order to verify these identities,
it is enough to show that the fusion rules dictated by
$S^\chi$ and the fusion rules dictated by composition of DHR-sectors
coincide.  Such identification of the two fusion rules have been
verified in many examples including all local conformal
nets with central charge less than 1 classified in \cite{KL}.
Also note that if these two representations of $SL(2,{\mathbb Z})$
coincide, we have the following Kac-Wakimoto formula, as explained
in  \cite[page 626]{FG}.
\begin{equation}\label{KW1}
d(\r)=\frac{S_{\r,0}}{S_{0,0}}=
\frac{S^\chi_{\r,0}}{S^\chi_{0,0}}=
\lim_{\tau\to i\infty}\frac{\sum_\nu S^\chi_{\r,\nu}\chi_\nu(\tau)}
{\sum_\nu S^\chi_{0,\nu}\chi_\nu(\tau)}=
\lim_{\tau\to 0}\frac{\chi_\r(\tau)}{\chi_0(\tau)}.
\end{equation}
Here we denote the vacuum sector by 0 and $d(\r)$ is the 
statistical dimension of $\r$.  (Note that we have $h_\r>0$ 
for $\r\neq0$, where $h_\r$ is the lowest eigenvalue of the 
operator $L_{0,\r}$, see Lemma \ref{hl}.)

We shall say that $\A$ is \emph{modular} if the $\mu$-index
$\mu_\A <\infty$ (see Sect.  \ref{mu-index}), the modular symmetries
\eqref{modularity} hold (in particular the characters are defined,
namely $\Tr(e^{-tL_{0,\r}})<\infty$) and the above two representations
of $SL(2,{\mathbb Z})$ are identical.  Note that a modular net is
completely rational.

Modularity holds in all computed rational case, cf. \cite{X}. 
The $SU(N)_k$ nets and the Virasoro nets ${\rm Vir}_c$ with $c<1$ are
both modular. We expect all local conformal completely rational 
nets to be modular (see \cite{Hu} for results of similar kind). 
Furthermore, we have the following.
\begin{proposition}
Let $\A$ be a modular local conformal net and $\B$ an irreducible extension 
of $\A$. Then $\B$ is also modular.
\end{proposition}
\proof
Since $\A$ is completely rational, the extension has finite index
and $\B$ is also completely rational.  We denote the $S$-matrices
for $\A$ and $\B$ arising from the braiding as in \cite{R1} by
$S$ and $\tilde S$, respectively.  For irreducible DHR sectors
$\r$ and $\sigma$ of $\A$ and $\B$, respectively, we put
$b_{\sigma,\r}={\dim}(\alpha_\r,\sigma)$, where
$\alpha_\r$ is $\alpha$-induction.  This $b_{\sigma,\r}$ is
equal to the multiplicity of $\r$ in the representation $\sigma$
restricted to $\A$.  Then we have
$\sum_{\sigma'} \tilde S_{\sigma,\sigma'} b_{\sigma',\r} =
\sum_{\r'}  b_{\sigma,\r'} S_{\r',\r}$ 
by \cite[Theorem 6.5]{BE4}.  We now have
\begin{eqnarray*}
\chi_\sigma(-1/\tau)&=&
\sum_\r b_{\sigma,\r} \chi_\r(-1/\tau)\\
&=&\sum_{\r,\r'} b_{\sigma,\r} S_{\r,\r'}
\chi_{\r'}(\tau)\\
&=&\sum_{\sigma',\r'} \tilde S_{\sigma,\sigma'} b_{\sigma',\r'} 
\chi_{\r'}(\tau)\\
&=&\sum_{\sigma'} \tilde S_{\sigma,\sigma'} \chi_{\sigma'}(\tau).
\end{eqnarray*}
This shows that the matrix $\tilde S$ arising from the braiding for
$\B$ also gives a transformation matrix for the characters.
\endproof
\begin{proposition}\label{K}
Assume that $\A$ is modular.  Then the following
asymptotic formula holds:
\[
\log\Tr(e^{-2\pi tL_0})\sim
\frac{\pi c}{12}\frac1t - \frac{1}{2}\log\mu_{\cal A}-\frac{\pi 
c}{12}t\qquad
{\rm as\ }t\to 0^+ \ .
\]
\end{proposition}
\proof
We first have
\[
\Tr(e^{-2\pi tL_0})= e^{-c\pi t/12}\sum_\nu
S_{0,\nu}e^{c\pi/(12t)}\Tr(e^{-2\pi L_{0,\nu}/t}).
\]
Then in this finite summation, the terms for $\nu\neq0$ are 
exponentially smaller than the term for $\nu=0$.  This gives
\[
\Tr(e^{-2\pi tL_0})\sim S_{00}e^{-\frac{\pi c}{12}(t- 1/t)}\ ,
\]
therefore
\[
\log\Tr(e^{-2\pi tL_0})\sim -\frac{c\pi }{12}t +
\log S_{00} +\frac{c\pi}{12}\frac1t,
\]
and we know that $S_{00}=\mu_{\cal A}^{-1/2}$ (e.g. \cite{R1}), so we get the
above statement.
\endproof
In particular, in the case $c<1$,
two-dimensional log-ellipticity can be proved for all 
local conformal nets. We give also an independent proof of this 
corollary as follows.
\begin{corollary}\label{ras}
Let $\A$ be a local conformal net with $c<1$. Then $\A$ is two-dimensional 
log-elliptic with noncommutative area $a_0 = 2\pi c/24$, thus 
\[
\log\Tr(e^{-2\pi tL_0}) \sim \frac{c}{24}\frac{2\pi}{t}
\quad {\rm as}\ \ t\to 0^+\ .
\]
\end{corollary}
\proof
The Virasoro net ${\rm Vir}_c$ with a central charge $c<1$ is
completely rational and $\A$ is a finite index extension of ${\rm Vir}_c$ 
\cite{KL}.  Hence the conformal Hamiltonian $L_0$ of $\A$ is a
finite direct sum of conformal Hamiltonians associated with
irreducible representations of ${\rm Vir}_c$.  As the stated
asymptotic is valid for all these conformal Hamiltonians \cite[Prop. 6.14]{W}, 
the proposition holds true.
\endproof
\begin{corollary}\label{Kr}
Let $\A$ be modular and $\r$ a representation of $\A$. 
The following asymptotic formula holds:
\[
\log\Tr(e^{-2\pi tL_{0,\r}})\sim \frac{\pi c}{12}\frac1t + 
\frac{1}{2}\log\frac{d(\r)^2}{\mu_{\A}}-
\frac{\pi c}{12}t\qquad {\rm as }\ t\to 0^+ \ .
\]
\end{corollary}
\proof 
We can assume $d(\r)<\infty$ as otherwise both members of the 
asymptotic equality are infinite.

By using Prop. \ref{K} and the Kac-Wakimoto 
formula (\ref{KW1},\ref{KW3}), we have
\begin{align*}
\log\Tr(e^{-2\pi tL_{0,\r}}) =&
\log\left(\Tr(e^{-2\pi tL_0})\frac{\Tr(e^{-2\pi tL_{0,\r}})}{\Tr(e^{-2\pi 
tL_0})}\right)\\
=& \log\Tr(e^{-2\pi tL_0}) + 
\log\frac{\Tr(e^{-2\pi tL_{0,\r}})}{\Tr(e^{-2\pi tL_0})}\\
\sim & \frac{\pi c}{12}\frac1t - 
\frac{1}{2}\log\mu_{\A}-\frac{\pi c}{12}t + \log d(\r)
\end{align*}
hence the corollary follows.
\endproof
We note explicitly that the information on the normalized index is 
contained in the spectral density of the Hamiltonian:
\[
\log d(\r) -\frac12\log\mu_\A = 
\lim_{t\to 0^+}\frac{\rm d}{{\rm d}t} t\log\Tr(e^{-tL_{0,\r}})\ .
\]
Because of the above formula we conjecure that a local confomal net $\A$ is 
completely rational iff 
\[
\lim_{t\to 0^+}\frac{\rm d}{{\rm d}t} t\log\Tr(e^{-tL_{0}})
<\infty.
\]
Recall now the following particular case of Kohlbecker's Tauberian theorem 
\cite[Th. 4.12.1]{BGT}. Let $m$ be a Borel measure on $[0,\infty)$ 
finite on compact sets.
The logarithm of the Laplace transform has the asymptotic 
behavior
\[
\log \int e^{-t\l}\text{d}m(\l)\sim C \frac1t \quad 
\text{as}\ t\to 0^+
\]
$C>0$, if and only if
\begin{equation}\label{Ko}
\log m[0,\l]\sim 2\sqrt{C\l},\quad \text{as}\ \l\to +\infty\ .
\end{equation}
As a further corollary, we then have an asymptotic formula which 
is, in part, a version of Cardy's formula \eqref{cardy}. Notice 
however that formula \eqref{cardy} concerns CFT on a 
two-dimensional spacetime, while we deal with conformal nets on $S^1$.
\begin{corollary}
Let $\A$ be a modular local conformal net on $S^1$ and $\r$ an 
irreducible representation of $\A$. Then
\[
\log N(\l)\sim 2\pi\sqrt{\frac{c}{6}\l}\quad \text{as}\ \l\to\infty
\]
where $N(\l)$ is the number of eigenvalues (with multiplicity) of 
$L_{0,\r}$ that are $\leq \l$.
\end{corollary}
\proof
By Cor. \ref{ras} we have $\log\Tr(e^{-t L_{0,\r}})\sim C/t$ with 
$C= \pi^2 c/6$. As
\[
\Tr(e^{-tL_{0,\r}})=\int e^{-t\l}{\rm d}m(\l)
\]
where $m[0,\l]=N(\l)$,
\eqref{Ko} reads   
$\log N(\l)\sim 2\sqrt{2\pi^2 c/12}\sqrt{\l}=2\pi\sqrt{c\l/6}$.
\endproof
From the physics viewpoint it is natural to define $S_\A$, the \emph{entropy 
of $\A$}, as the leading coefficient of the expansion \eqref{asymp} of 
$\log\Tr(e^{-2\pi t L_0})$, thus
\begin{align*}
   & a_0 = S_\A \ , \\
   & a_1, a_2, \dots = \text{higher order corrections to}\ S_\A \ .
\end{align*}
Note that, by definition, the entropy is proportional to the noncommutative 
area: it is just a matter of reading the same formula from different 
point of views.
\subsection{The incremental free energy in \cite{L2} (increment of 
the first spectral invariant)}\label{dF}
Let $\A$ be a local conformal net and $\r$, $\s$ a DHR representation of 
$\A$ (see Sect.  \ref{nets}). The above mentioned Kac-Wakimoto formula 
\begin{equation}\label{KW3}
\lim_{t\to 0^+}
\frac{\Tr(e^{-t L_{0,\r}})}{\Tr(e^{-t L_{0,\s}})}
= \frac{d(\r)}{d(\s)}\ .
\end{equation}
has been tested in wide generality and always holds true, see \cite{X}, 
and we have just seen to hold true if $\A$ is modular.
\begin{proposition}\label{qekw}
If $\A$ is modular, then
\[
\log d(\r) - \log d(\s) =a_1(2\pi L_{0,\r}) - 
a_1(2\pi L_{0,\s})
\equiv \tfrac{1}{12}(\chi_{\r}-\chi_{\s}) 
\]
where $\chi_{\r}-\chi_{\s}$ is 
the increment of the noncommutative Euler characteristic by adding the 
charge $\r$ and removing the charge $\s$. 
\end{proposition}
\proof This is an immediate corollary of Prop. \ref{Kr}.
\endproof
Recall now the work in \cite{GLRV,L3} in the context of black holes 
described by a curved spacetime with a bifurcate Killing horizon.  
There $\A$ is a local conformal net canonically arising on the 
horizon. According to the general analysis (by using Wiesbrock's 
theorem) $\A$ is a M\"{o}bius covariant net, but $\A$ is expected to 
be diffeomorphism covariant too \cite{We}; for example this is the case when the 
quantum field is free, as $\A$ is then isomorphic to the net 
associated with the $U(1)$-current algebra \cite{GLRV} (see also \cite{MP}).

The incremental free energy ${\rm d}F$ by adding the charge $\r$ and 
removing the charge $\s$ (in the Hartle-Hawking state) in \cite{L2} 
or, more generally, its symmetrization, see \cite[Thm. 5.4]{L4}, is 
defined and turns out to be given by
\begin{equation}\label{dF2}
{\rm d}F = \b\big(\log d(\r) - \log d(\s)\big) = 
\frac{2\pi}{\kappa}\big(\log d(\r) - \log d(\s)\big) 
\end{equation}
where $\kappa$ is the surface gravity and 
$\b\equiv 2\pi/\kappa$ is the Hawking temperature.

We thus assume $\A$ to be diffeomorphism covariant and that Prop. 
\ref{qekw} holds. Recall that, in higher dimensional spacetimes, 
$d(\r)\in\mathbb N\cup\infty$ \cite{DHR}. We then have:
\begin{corollary}
With the above assumptions, the incremental free energy by adding 
the DHR charge $\r$ and removing the charge $\s$
is proportional to the increment of the noncommutative Euler 
characteristic 
\begin{equation}\label{dF3}
{\rm d}F = \frac{\pi}{6\kappa}\big(\chi_{\s}-\chi_{\r}\big). 
\end{equation}
Adding a charge is proportional to the logarithm of an integer. 
\end{corollary}
\proof
The proof is immediate from the above discussion.
\endproof

The above formulas (\ref{dF2},\ref{dF3}) are consistent with the 
interpretation of the entropy by logarithmic counting states and the 
fact that it is proportional to an integer as in eq.  
(\ref{discr.entr}).

Compared with the work \cite{L2}, the above corollary expresses the 
incremental free energy by a true difference of global entropies 
$\log\Tr(e^{-tL_{0,\r}})$ and $\log\Tr(e^{-tL_{0,\s}})$ by Prop. \ref{qekw}.

\subsection{Relation to black hole entropy. I}
\label{BKE1}
A microscopic derivation of black hole entropy and its relation to 
conformal symmetries and central charge is discussed in \cite{SV}.  
The potentiality of our discussion in relation to black hole entropy 
and Bekenstein classical area description is well exemplified if one 
relies on the reference \cite{Ca} recalled in Sect.  \ref{BKE}.  Yet we use 
here only the value of the central charge (eq.  \eqref{Carlip}) and 
not Cardy's formula nor the boundary term of the energy.  We shall make 
here the assumption that the associated local conformal net $\A$ is 
modular.  (Later we shall introduce the mean free energy and put it in 
relation to Bekenstein entropy, on the same lines, 
without the modularity assumption.)  
\begin{corollary}\label{area}
For a black hole in the above class \cite{Ca}, we have
\[
S_\A = A/4
\]
where $A$ is the area of the black hole horizon.
\end{corollary}
\proof
Immediate from the relation $c/12= A/8\pi$ 
\eqref{Carlip} and the value $S_\A= 2\pi c/12$ of the  
entropy for modular nets on the two-dimensional Minkowski spacetime.
\endproof
We have therefore the picture in the following diagram:
\[
\CD
\text{Entropy} @>\text{physics}>>
a_0  @<\text{geometry}<< 4\pi\cdot\text{Noncommutative area}
\\ & &  @ V \text{modular} V \text{nets} V 
&  \\  
& & 2\pi c/12 & \\
& &  @ V \text{black hole} V \text{models} V \\
& & A/4 &
\endCD
\]
\section{Discretization and conformal invariance}
There is an apparent conflict in regarding the horizon of a black hole 
both having a discrete essence and a conformal group of symmetries. In 
the sequel we take simultaneously account of both pictures by considering the 
$n$-cover $\Diff^{(n)}(S^1)$ of $\Diff(S^1)$ acting on $S^1$ and respecting the cell 
partitioning of $S^1$. Thus the conformal Hamiltonian becomes the 
generator of the rotation group for the unitary action of $\Diff^{(n)}(S^1)$.
We then consider mean quantities, as entropy, as $n$ tends to infinity. 

Note that in the sequel of this paper we shall not any longer need 
the modularity or log-ellipticity assumptions.
\subsection{The action of the $n$-cover of $\Diff(S^1)$} We recall now 
some facts on $\Diff^{(n)}(S^1)$ and its canonical embedding into 
$\Diff(S^1)$, see \cite{LX}.

The Virasoro algebra is the infinite dimensional Lie algebra 
generated 
by elements $\{L_n \mid n\in\mathbb Z\}$ and $c$ with
relations
\begin{equation}\label{vir-rel}
[L_m,L_n]=(m-n) L_{m+n} + \frac{c}{12}(m^3-m)\delta_{m,-n}.
\end{equation}
and $[L_n,c]=0$. It is the (complexification of) the
unique, non-trivial one-dimensional central extension of the Lie 
algebra 
of $\Vect(S^1)$.

The elements $L_{-1},L_{0},L_{1}$ of the Virasoro 
algebra are clearly a basis of $s\ell(2,\mathbb C)$.
The Virasoro algebra  contains infinitely many
further copies of $s\ell(2,\mathbb C)$: for every fixed $n>0$ we get a 
copy generated by the elements
$L^{(n)}_{-1},L^{(n)}_{0},L^{(n)}_{1}$, where
\begin{gather}
L^{(n)}_{\pm 1} \equiv \frac{1}{n}L_{\pm n}\ , \\
L^{(n)}_{0} \equiv \frac{1}{n}L_{0} +  \frac{c}{24}\frac{(n^{2}-1)}{n}\ .
\label{spin}
\end{gather}         
We have indeed
\begin{equation}
  [L^{(n)}_1,L^{(n)}_{-1}]=2L^{(n)}_{0},\qquad
         [L^{(n)}_{\pm 1},L^{(n)}_{0}]= \pm L^{(n)}_{\pm 1}
\end{equation}
that are the relations for the usual generators in 
$s\ell(2,\mathbb C)$.

It follows that, setting for a fixed $n>0$
\begin{equation}
L^{(n)}_m \equiv \frac{1}{n}L_{nm}\ , \quad m\neq 0\ ,
\end{equation}  
and $L^{(n)}_0$ as in \eqref{spin}, the map 
\begin{equation*}
\left\{\begin{array}{l}    
 L_m\mapsto  L^{(n)}_m\\
 c\mapsto nc\ ,
\end{array}\right.
\end{equation*}
gives an embedding of the Virasoro algebra into itself. 
There corresponds an embedding of 
$\Diff^{(n)}(S^1)$, the $n$-cover of $\Diff(S^1)$, into $\Diff(S^1)$ 
as stated in the following.
\begin{proposition}\cite{LX}\label{cover2}
There is a unique continuous isomorphism $M^{(n)}$ of 
$\Diff^{(n)}(S^1)$ into 
$\Diff(S^1)$ such that for all $g\in\Diff^{(n)}(S^1)$ the following diagram commutes
\begin{equation}\label{cdn}
\begin{CD}
S^1 @>M^{(n)}_g >> S^1   \\
@V z^n V V @V V z^n V\\
S^1 @>M_{\underline g}>> S^1
\end{CD}
\end{equation}
i.e. $M_g^{(n)}(z)^n=M_{\underline g}(z^n)$ for all $z\in S^1$.

Here ${\underline g}$ is the element of $\Diff(S^1)$ corresponding to 
$g$ and $M_{\underline g}$ is the obvious action of ${\underline g}$ on $S^1$.

\Mob$^{(n)}\equiv\{g\in \Diff^{(n)}(S^1):\ {\underline g}\in 
\text{\Mob}\}$ is the $n$-cover of {\Mob} 
and $M^{(n)}$ restricts to an embedding of \Mob$^{(n)}$ into $\Diff(S^1)$.
\end{proposition}
\noindent Clearly the embedding \Mob$^{(n)}\hookrightarrow\Diff(S^1)$ 
corresponds to the embedding $L_m\mapsto L^{(n)}_{m}$, $m=-1,0,1$, of 
$s\ell(2,\mathbb C)$ into the Virasoro algebra.
\subsection{The mean free energy (topological increment of the second 
spectral invariant)}

Let $\A$ be a local conformal net on $S^1$ (in any representation).
We divide $S^1$ into $n$ equally spaced cells, namely we consider the
$n$-interval $E_n\equiv\sqrt[n]{S^+}$, where $S^+$ is the upper
semicircle.  Each interval component $I_k$ of $E_n$ contains minimal
information (as the cells of Planck length).  There is a canonical
evolution associated with $E_n$ corresponding to the rotations on the
full $S^1$, namely the rescaled rotations $R(\frac{1}{n}\vartheta)$,
giving rise of two rescaled conformal Hamiltonians: one, ${\hat
L}^{(n)}_0\equiv \frac{1}{n}L_0$, comes by purely rescaling the
Hamiltonian, the other is the one associated with the representation
$U^{(n)}$ of $\Diff^{(n)}(S^1)$, namely $L^{(n)}_0=\frac{1}{n}L_0 +
\frac{c}{24}\frac{(n^2 - 1)}{n}$, and takes care of ``boundary effects''.
The geometrical complexity should be encoded in the difference between
the two terms.

We define the free energy associated 
with the above partition of $S^1$ as the difference of the free energy 
associated by the corresponding partition functions at infinite 
temperature:
\[
F_n\equiv t^{-1}\log\Tr(e^{-t2\pi{L}^{(n)}_0}) - 
t^{-1}\log\Tr(e^{-t2\pi{\hat L}^{(n)}_0})\ .
\]
(one could generalize the definition of $F_n$ without the existence of 
characters, but we do not dwell on this point).
Clearly $F_n = \frac{c}{24}\frac{(n^2 - 1)}{n}2\pi$ hence we 
get the following model independent formula for the mean free energy 
associated to the ``discretization of $S^1$''.
\begin{theorem}\label{topentropy}
let $\A$ be a local conformal net. We have
\begin{equation}
	\label{Fmean}
F_{\rm mean} = 2\pi\frac{c}{24}
\end{equation}
\end{theorem}
\proof 
Obviously 
$F_{\rm mean}\equiv \lim_{n\to\infty}\frac{1}{n} F_n = 2\pi c/24$.  
\endproof 
Note that we clearly have the relation
\[
a_2(2\pi{L}^{(n)}_0) - a_2(2\pi{\hat L}^{(n)}_0) = F_n
\]
thus also $F_{\rm mean}$ has a noncommutative geometrical meaning.

Concerning a two-dimensional conformal QFT, both chiral components 
contribute to the topological entropy thus, assuming the central 
charge to be equal for both components, the physical topological 
entropy duplicates:
\begin{equation}\label{2F}
F_{\rm mean}=2\pi\frac{c}{12}\ ,
\end{equation}
we shall explain this point in Sect. \ref{CFT2}.
\subsection{Relation to black hole entropy. II}
As noted, the derivation of the value $F_{\rm mean} = 2\pi c/12$ is model 
independent and general, essentially it follows only by diffeomorphism 
invariance. 

As the value of $F_{\rm mean}$ coincides with the value of $S_\A$ 
(for modular nets), we now have a link with the classical area 
restriction, just as in Sect. \ref{BKE1}, without any modularity
assumption on $\A$.

For a black hole as in the Corollary \ref{area}, we have indeed
\[
F_{\rm mean}= A/4
\]
where $A$ is the area of the black hole horizon.
This is immediate from the relation $c/12 = A/8\pi$
\eqref{Carlip} and the found value $F_{\rm mean}= 2\pi c/12$ of
the two-dimensional free energy \eqref{2F}.  
\section{The modular group of a $n$-interval von Neumann algebra}
\label{modulargroup}

Here we extend to the general model independent setting, and in an
arbitrary representation, a formula (announced in \cite{LX}) for the
modular group discussed by Schroer and Wiesbrock \cite{SW} in the
context of the $U(1)$-current algebra local conformal net.

Let $E$ be a symmetric $n$-interval of $S^1$, thus 
$E\equiv\sqrt[n]{I}$ for some $I\in\I$, i.e. $E=\{z\in S^1 :z^n\in I\}$.
Let $I_0 , I_1, \cdots I_{n-1}$ be
the $n$ connected components of $E$; we may assume that $I_k = R(2\pi 
k/n)I_0$, where $R$ is the rotation subgroup of $\Mob$. 

Let $\A$ be a local conformal net on $S^1$ with the 
split property, in a irreducible representation.
By the split property we have 
a natural isomorphism
\[
\chi_E:\A(E)\equiv\A(I_0)\vee\A(I_1)\vee\cdots\vee\A(I_{n-1})\to 
\A(I_0)\otimes\A(I_1)\otimes\cdots\otimes\A(I_{n-1}) \ .
\]
A product state $\f$ is a state on $\A(E)$ of the form
\[
\f\equiv (\f_0\otimes\f_1\otimes\cdots\otimes\f_{n-1})\cdot\chi_E\ ,
\]
where $\f_k$ is a normal faithful state on $\A(I_k)$ and 
$\f_k=\f_0\cdot \Ad U(R(2k\pi/n))$ is called  a \emph{rotation 
invariant product state}. 

We now exhibit a modular group of $\A(E)$ having a geometrical meaning.

Let $\Phi_k$ be the isomorphism between $\A(I_k)$ and $\A(I)$ associated 
with the function ${z^n}$, namely
\[
\Phi_k(x)\equiv U(h_k)xU(h_k)^*,\quad x\in \A(I_k)
\]
where $h_k$ is any element of $\Diff(S^1)$ such that $h_k(z)  = z^n$, 
$z\in I_k$, (by locality the definition of $\Phi_k$ is independent of 
the choice of $h_k$).

Let $\f_k$ be the state on $\A(I_k)$ given by 
$\f_k\equiv \o_I\cdot \Phi_k$, where $\o$ is the vacuum state, 
and let $\f_E$ the product state on $\A(E)$ that restricts 
to $\f_k$ on $\A(I_k)$. Clearly $\f_E$ is a rotation invariant product 
state.
\begin{theorem}
\label{modgroup}
Let $\A$ be a local conformal net in a irreducible representation and 
$U$ the covariance unitary representation of $\Diff(S^1)$.  With 
$E=\sqrt[n]{I}$ an $n$-interval as above, the canonical rotation 
invariant product state $\f_E$ on $\A(E)$ has modular group 
$\s^{\f_E}$ given by
\[
\s_t^{\f_E}=\Ad U^{(n)}(\Lambda_I(-2\pi t))\res_{\A(E)}
\]
where $\Lambda_I$ is the lift to \Mob$^{(n)}$ of the one-parameter 
subgroup of {\rm \Mob} of generalized dilatation associated with $I$ 
(see Appendix \ref{nets}) and $U^{(n)}=U\cdot M^{(n)}$ is the unitary 
representation of \Mob$^{(n)}$ associated with $U$.
\end{theorem}
\proof 
Since both $\s_t^{\f_E}$ and 
$\Ad U^{(n)}(\Lambda_I(-2\pi t))\res_{\A(E)}$
are tensor product of their restrictions to the components 
$\A(I_k)$, by rotation invariance it suffices to prove the formula on 
each $\A(I_k)$.

We have
\begin{multline}
\s_t^{\f_E}\res_{\A(I_k)}=\s_t^{\o\cdot \Phi_k}\res_{\A(I_k)}=
\Phi_k^{-1} \cdot \s_t^{\o_I}\cdot \Phi_k\\
=\Phi_k ^{-1} \cdot \Ad U(\Lambda_I(-2\pi t))\restriction_{\A(I)}\cdot 
\Phi_k
= \Ad U^{(n)}(\Lambda_I(-2\pi t))\restriction_{\A(I_k)}\ .
\end{multline}
\endproof
\begin{corollary}\label{V}
In the above proposition, setting 
$V(t)\equiv U^{(n)}(\Lambda_I(-2\pi t))$, 
we have:
\[
\Ad V(t)\restriction_{\A(E)} = \s_t^{\f_E},
\qquad \Ad V(-t)\restriction_{\A(E')} = \s_t^{\f_{E'}}\ .
\]
\end{corollary}
\proof 
The first equality has been already shown. Since $E' = \sqrt[n]{I'}$, 
to get the second equality we just have to show that $V(-t) = 
U^{(n)}(\Lambda_{I'}(-2\pi t))$, which is clearly the case since
$\Lambda_{I'}(-t)= \Lambda_{I}(t)$.
\endproof
Note that the abstract results in Appendix \ref{math} now apply.
\section{Entropy and global index with the proper Hamiltonian}

In this section we pursue the above point of view, but we 
replace the conformal Hamiltonian $L_0$  with the ``local'' Hamiltonian 
\[
K_1\equiv i(L_1 - L_{-1})\ ,
\]
the generator of the one-parameter dilatation 
unitary group associated with the upper semicircle $S^+$ (see 
Appendix \ref{nets}). With this dynamics, the restriction of the 
vacuum state satisfies the equilibrium condition at 
Hawking temperature and is natural to be considered, see e.g. 
\cite{H,Wa,L2,L4}. As above we will consider the corresponding 
dynamics for the action of $\Diff^{(n)}(S^1)$ and compute 
noncommutative spectral invariants. It turns out the analysis below 
can be done in complete generality: it is only based on conformal 
invariance ad the split property (recall that the latter follows 
automatically from the existence of characters). 
\subsection{$\mu$-index}
\label{mu-index}

Let $\A$ be a local conformal net with the split property in the 
vacuum representation and $E\subset S^1$ a 
2-interval, namely $E$ and its complement $E'$ are union of two 
proper intervals. The \emph{$\mu$-index} of $\A$ is defined as
\[
\mu_\A \equiv [\hat\A(E):\A(E)]
\]
where the brackets denote the Jones index and $\hat\A(E)\equiv\A(E')'$. 
It turns out that $\mu_{\A}$ does not depends on $E$ and 
\[
\mu_{\A}=\sum_i d(\r_i)^2
\]
sum over the indices of all irreducible DHR charges, namely $\mu_\A$ 
coincides with the \emph{global index} of $\A$.  More generally, if 
$E_n$ is an $n$-interval, and in the representation $\r$, we have
\[
\mu^{\r}_{\A,n}\equiv [\hat\A(E_n):\A(E_n)]=d(\r)^2\mu_{\A}^{n-1}\ .
\]
Note that the formula 
\[
\mu_{\A} = \lim_{n\to\infty}\sqrt[n]{[\hat\A(E_n):\A(E_n)]}
\]
gives the $\mu$-index in any irreducible representation. Indeed we 
have:
\begin{proposition}
Let $\A$ be a split, local M\"{o}bius covariant net in a
irreducible representation $\r$. 
Given an interval $I$, both $\mu_{\A}$ and 
$d(\r)$ can be measured in $I$.
\end{proposition}
\proof
Fix be an interval $I$ and divide $I$ in $2n -1$ contiguous intervals $I_1 < 
J_1 < I_2 < J_2 < \cdots < J_{n-1} < I_n$, where $<$ denotes the 
counter-clockwise order. Then $\vee_{i=1}^{n}\A(I_i)\subset 
\big(\vee_{i=1}^{n-1}\A(J_i)\big)'\cap\A(I)$
is an $n$-interval inclusion, thus its index is equal to 
$d(\r)^2\mu_{\A}^{n-1}$ and we have
\[
\lim_{n\to\infty}[\big(\vee_{i=1}^{n-1}\A(J_i)\big)'\cap\A(I):
\vee_{i=1}^{n}\A(I_i)]^{\frac1n}= 
\lim_{n\to\infty}\big(d(\r)^2\mu^{n-1}_{\A}\big)^{\frac{1}{n}}=\mu_\A
\]
showing that $\mu_{\A}$ can be detected within the interval $I$ and so is the case 
also for 
$d(\r)^2= 
[\big(\vee_{i=1}^{n-1}\A(J_i)\big)'\cap\A(I):
\vee_{i=1}^{n}\A(I_i)]/\mu_{\A}^{n-1}$ 
(for instance with $n=2$).
\endproof
As is known, the central charge may also be measured locally, as it 
appears locally in the commutation relations with the stress-energy 
tensor.
\subsection{$\mu$-entropy and spectral invariants for the proper 
Hamiltonian}
Let $\A$ be a local conformal net on $S^1$ with the split property in 
an irreducible representation $\r$. Let $I= S^+$ be 
the upper semicircle, $E\equiv E_n=\sqrt[n]{I}$ the associated $n$-interval 
and  $K_n$ the infinitesimal generator of $V^{(n)}$, where 
$V^{(n)}(t)=U^{(n)}(\Lambda_I(-2\pi t))$ as in Cor. \ref{V}.
Note that
\[
K_n \equiv i(L^{(n)}_1 -L^{(n)}_{-1}) = \tfrac{i}{n}(L_n -L_{-n})\ .
\]
The complement $E'_n$ of $E_n$ is the $n$-interval $E'_n=\sqrt[n]{I'}$.
Let $\f_{E_n}$ be the  rotation-invariant product state on 
$\A(E_n)$ defined in Prop. \ref{modgroup}
and $\xi_n\equiv\xi_{E_n}$ a cyclic separating vector for $\A(E_n)$ 
implementing $\f_{E_n}$. 
\begin{theorem}\label{lkw}
 We have
\[
(e^{-2\pi K_n}\xi_n,\xi_n) = d(\r)\mu_{\A}^{\frac{n-1}{2}} \ ,
\]
thus 
\[
\log (e^{-\frac{2\pi i}{n}(L_n - L_{-n})}\xi_n,\xi_n) = 
\tfrac{n-1}{2}\log \mu_{\A} + \log d(\r) = 
\tfrac{n-1}{2}\log(\sum_i d(\rho_i)^2) +\log d(\r)
\]
\end{theorem}
\proof 
The unitary $U(R(2\pi/n))$ implements an isomorphism between
$\A(E_n)$ and $\A(E'_n)$, and between $\hat\A(E_n)$ and
$\hat\A(E'_n)$; moreover it maps $\f_{E'_n}$ to $\f_{E_n}$ and $K_n$
to $-K_n$.  Hence, if $\xi'_n$ is a cyclic and separating vector for
$\A(E'_n)$ implementing the state $\f_{E'_n}$, we have $(e^{-2\pi
K_n}\xi_n,\xi_n)=(e^{2\pi K_n}\xi'_n,\xi'_n)$, thus by Cor.
\ref{absf}
\[
(e^{-2\pi K_n}\xi_n,\xi_n)^2 = \mu^{\r}_{\A,n}\equiv
[\hat\A(E_n):\A(E_n)]=d(\r)^2\mu_{\A}^{n-1}\ .
\]
\endproof
If the $\mu$-index is finite, we shall denote by
\[
\hat\f_{E_n} = \f_{E_n}\cdot\e_{E_n}
\]
the state on $\hat\A(E_n)$ obtained by extending $\f_{E_n}$ by 
the conditional expectation $\e_{E_n}:\hat\A(E_n)\to\A(E_n)$. The 
state $\hat\f_{E'_n}$ on $\hat\A(E'_n)$ is defined analogously. If 
$\mu_{\A}=\infty$ there exists an operator-valued weight 
$\e_{E_n}:\hat\A(E_n)\to\A(E_n)$ by Haagerup theorem and Prop. \ref{modgroup}, 
but for our purposes here we can stay in the finite $\mu$-index case. 
\begin{corollary}
We have
\begin{align*}
K_n \equiv \tfrac{i}{n}(L_n -L_{-n})
&= -\frac{1}{2\pi}\Big(\log\big(\frac{{\rm d}\f_{E_n}}{{\rm 
d}\hat\f_{E'_n}}\big)  + \tfrac{n-1}{2}\log\mu_\A + \log d(\r)\Big)\\
&=  -\frac{1}{2\pi}\Big(\log\big(\frac{{\rm d}\hat\f_{E_n}}{{\rm 
d}\f_{E'_n}}\big)  + \tfrac{n-1}{2}\log\mu_\A + \log d(\r)\Big) 
 \ .
\end{align*}
\end{corollary}
\proof The von Neumann algebra $\A(I)$ associated to an interval is a
factor \cite{BGL} hence, by the spit property, also the von Neumann
algebra $\A({E_n})$ associated with the $n$-interval ${E_n}$ is a
factor.  As both $V^{(n)}(t)$ and $\big(\frac{{\rm
d}\hat\f_{E_n}}{{\rm d}\f_{E'_n}}\big)^{it}$ implement
$\s_t^{\hat\f_{E_n}}$ on $\A(E_n)$ and $\s_{-t}^{\f_{E'_n}}$ on
$\hat\A(E_n)$, we have that $-2\pi K_n$ is equal to $\log({\rm
d}\hat\f_{E_n}/{\rm d}\f_{E'_n})$ plus a constant term (see 
Appendix \ref{math}).  Such constant
is fixed by Th.  \ref{lkw} to be $\tfrac{n-1}{2}\log\mu_\A +\log
d(\r)$.  \endproof
The quantity 
\[
Z_n(t)\equiv (e^{-t K_n}\xi_n,\xi_n)
\]
is the geometric partition function associated to the symmetric
$n$-interval partition of $S^1$, thus by Th.  \ref{lkw}
\begin{equation}\label{sen}
F_{n,\mu} \equiv -t^{-1}\log Z_n(t)|_{t=2\pi} = -\tfrac{n-1}{4\pi}\log\mu_{\A}
-\tfrac{1}{2\pi}\log d(\r)	
\end{equation}
is the associated $n$-free energy, that we call the 
\emph{$n$-$\mu$-free energy}. The $n$-$\mu$-free energy divided by the 
numbers of cells (intervals) gives asymptotically the \emph{mean $\mu$-free energy}.
\begin{corollary}
The mean $\mu$-free energy is given by
\[
F_{{\rm mean},\mu} = -\tfrac{1}{4\pi}\log\mu_{\A}\ .
\]
\end{corollary}
\proof
Immediate by eq. \eqref{sen} we have
\begin{multline}\label{mue}
F_{{\rm mean},\mu}\equiv
\lim_{n\to\infty}\tfrac{1}{n}F_{n,\mu}=
-\lim_{n\to\infty}\tfrac{1}{2\pi n}\log (e^{-2\pi K_n}\xi_n,\xi_n)\\
=-\lim_{n\to\infty}\big(\tfrac{n-1}{4\pi n}\log\mu_{\A} 
+\tfrac{1}{2\pi n}\log d(\r)\big)=
-\tfrac{1}{4\pi}\log\mu_{\A}\ .
\end{multline}

\endproof
In analogy with Sect. \ref{si}  the \emph{ $0^{\rm th}$ and $1^{\rm 
st}$ spectral invariants} are then defined by
\begin{align}
a_{0,\mu} & \equiv 
\lim_{n\to\infty}\frac{t\log Z_n(t)}{n}\arrowvert_{t=2\pi} \\
a_{1,\mu} & \equiv 
\lim_{n\to\infty}\frac{\rm d}{{\rm d}t}\frac{t\log Z_n(t)}{n}\arrowvert_{t=2\pi}
\end{align}
Note that $-\frac{\rm d}{{\rm d}t}\log Z_n(t)$ is the  
$n-\mu$-\emph{energy} $H_{n,\mu}$ associated with $Z_n(t)$. Due to the 
thermodynamical relation
\[
\text{Free energy} = T\cdot\text{Entropy} - \text{Energy}
\]
where $T$ is the  temperature, we thus define   
the mean $n-\mu$-\emph{entropy} by
\[
S_{n,\mu} = t(F_{n,\mu} + H_{n,\mu}) \ .
\]
We have:
\begin{align}
\frac{\rm d}{{\rm d}t}t\log Z_n(t) &= \log Z_n(t) + t\frac{\tfrac{\rm d}{{\rm 
d}t}Z_n(t)}{Z_n(t)}\\
&= -t\big(F_{n,\mu} + H_{n,\mu})\\
&= -S_{n,\mu}
\end{align}
thus the mean $\mu$-\emph{entropy} at 
Hawking inverse temperature $2\pi$ is given by
\[
S_{{\rm mean},\mu}=\lim_{n\to\infty} S_{n,\mu}/n
=-\lim_{n\to\infty}\frac{\rm d}{{\rm d}t}\frac{t\log 
Z_n(t)}{n}\arrowvert_{t=2\pi}\ .
\] 
\begin{proposition}
$S_{n,\mu}=S(\hat\f_{E_n}|\f_{E'_n})$, where the latter is Araki relative entropy 
between the states $\hat\f_{E_n}$ and $\f_{E'_n}$.
\end{proposition}
\proof
We fix a natural cone $L^2(\A(E_n))_+$ (that is unique up to unitary 
equivalence); for example, in the vacuum representation, we can take 
the natural cone with respect to the vacuum vector $\Omega$.

The derivative of $\log Z_n(t)$ at $t=2\pi$ is given by
\begin{align*}
\frac{\rm d}{{\rm d}t}\log(e^{-tK_n}\xi_n,\xi_n)\arrowvert_{t=2\pi}=&
-\frac{(K_n e^{-tK_n}\xi_n,\xi_n)}{(e^{-tK_n}\xi_n,\xi_n)}\big\arrowvert_{t=2\pi}
\\=& -\mu_{\A,n}^{\r}
\frac{(K_n\Delta^{1/2}\xi_n,\Delta^{1/2}\xi_n)}
{(e^{-tK_n}\xi_n,\xi_n)}\big\arrowvert_{t=2\pi}
\\=& -(K_n JJ\Delta^{1/2}\xi_n,JJ\Delta^{1/2}\xi_n)
\\=& -(K_n\hat\xi'_n,\hat\xi'_n)
\\=& \tfrac{1}{2\pi}\big( (\log\Delta\hat\xi'_n,\hat\xi'_n) + 
\tfrac12\log\mu_{\A,n}^{\r}\big)
\\=& t^{-1}\big(-S(\hat\f_{E_n}|\f_{E'_n}) 
+ \tfrac12\log\mu_{\A,n}^{\r}\big)\arrowvert_{t=2\pi}
\end{align*}
where $\Delta\equiv\Delta_{\hat\xi'_n , \xi_n}$ is Araki relative
modular operator between the vectors $\xi_n, \hat\xi'_n\in L^2(\A(E_n))_+$
implementing the states $\f_{E_n}$
on $\A(E_n)$ and $\hat\f_{E'_n}$ on $\hat\A(E'_n)$, and $J$ is the
corresponding modular conjugation.  Hence
\begin{align*}
\frac{\rm d}{{\rm d}t}t\log(e^{-tK_n}\xi_n,\xi_n)\arrowvert_{t=2\pi}
=& \log(e^{-tK_n}\xi_n,\xi_n)\arrowvert_{t=2\pi} 
+t \frac{\rm d}{{\rm d}t}\log(e^{-tK_n}\xi_n,\xi_n)\arrowvert_{t=2\pi}
\\=& \log(e^{-tK_n}\xi_n,\xi_n)\arrowvert_{t=2\pi}
 -  S(\hat\f_{E_n}|\f_{E'_n}) + \tfrac12\log\mu_{\A,n}^{\r}
\\=&  \tfrac{1}{2}\log\mu_{\A,n}^{\r} -  
S(\hat\f_{E_n}|\f_{E'_n}) + \tfrac{1}{2}\log\mu_{\A,n}^{\r}
\\=& - S(\hat\f_{E_n}|\f_{E'_n}) + \log\mu_{\A,n}^{\r}\\
=& - S(\hat\f_{E_n}|\f_{E'_n}) + (n-1)\log\mu_{\A} + \log d(\r)
\end{align*}
which gives the thesis.
\endproof
\begin{corollary}
We have
\begin{align*}
a_{0,\mu}=& \tfrac12\log\mu_{\A}\ ,\\
a_{1,\mu}=& -S_{{\rm mean},\mu} = \log\mu_\A - 
\lim_{n\to\infty}\tfrac{1}{n} S(\hat\f_{E_n}|\f_{E'_n}) \ .
\end{align*}
\end{corollary}
\proof
Immediate by the above discussion.
\endproof
By definition the $\mu$-\emph{noncommutative Euler characteristic} $\chi_{\A,\mu}$ 
is defined, in analogy with the previous sections, to be equal to 12 
times the first spectral invariant. Thus we have: 
\[
\chi_{\A,\mu}\equiv 12 a_{1,\mu}=-12 S_{{\rm 
mean},\mu} \ .
\]
\section{CFT on a two-dimensional spacetime}
\label{CFT2}
Here we give the version of the considered asymptotic expansion in 
the case of a conformal QFT on a two-dimensional spacetime. The 
extension of the rest of our analysis is then immediate and we do not 
make it explicitly.

The model independent structure of  conformal quantum 
field theory on the two-dimensional Minkowski spacetime $M_2$ is naturally described 
by a local, diffeomorphism covariant net $\A$ of von Neumann 
algebras $\A({\cal O})$ associated with double cones ${\cal O}$ of $M_2$, see e.g. 
\cite{KL2}.

Denoting with $(x,t)$ the space and time coordinates of a point of $M_2$, 
the restriction of $\A$ to the light axis $x\pm t = 0$ gives rise to 
two local chiral conformal nets $\A_{\pm}$ on $\mathbb R$ that, by 
conformal invariance, extend to local conformal nets on $S^1$.

Given the double cone
\[
{\cal O}
=\{(x,t): x\pm t \in I_{\pm}\}
\]
associated with the intervals $I_+$ and $I_-$ of the light axis, denote 
by $\A_0({\cal O})$ the von Neumann algebra
\[
\A_0({\cal O}) = \A_+(I_+)\vee\A(I_-)\simeq \A_+(I_+)\otimes\A(I_-) ;
\]
then $\A_0$ is a local conformal subnet of $\A$. In the rational case 
one expects the subnet to have finite Jones index:
\[
[\A({\cal O}):\A_0({\cal O})]<\infty \ .
\]
This is the case if $\A_0$ is completely rational, namely if
$\A_{\pm}$ are completely rational, which is automatic for example if
the central charge(s) of $\A$ (i.e. of $\A_{\pm})$ are less than one.

The classification of all local conformal nets on $M_2$ with central 
charge $c<1$ has been obtained in \cite{KL2}.

We shall say that $\A$ is \emph{modular} if both $\A_+$ and $\A_-$ are 
modular.

Rehren describes the structure of the inclusion $\A_0({\cal O})\subset\A({\cal O})$ 
in terms of modular invariants \cite{R4}.

The restriction to $\A_0$ of the identity representation of $\A$ 
decomposes as $\bigoplus Z_{ij} \r^+_{i}\otimes\r^-_{j}$ 
with $\{\r^{+}_i\}$ and $\{\r^{-}_i\}$ irreducible sectors of $\A_+$ and $\A_-$.
Accordingly, the conformal Hamiltonian $H$ of $\A$ (the generator of the 
rotation one-parameter group in the time direction), has a 
decomposition
\[
e^{-tH} = \bigoplus_{i,j} Z_{ij} e^{-tL^+_{0,i}}\otimes e^{-tL^-_{0,j}}\
\]
where $L^{\pm}_{0,i}$ is the conformal Hamiltonian of $\A_{\pm}$ in the 
representation $\r^{\pm}_i$.
\begin{proposition}\label{2as}
Let $\A$ be a modular local conformal net on the two-dimensional 
Minkowski spacetime.
We have the expansion as $t\to 0^+$:
\[
\log\Tr(e^{-2\pi tH})\sim
\frac{2\pi c}{12}\frac1t - \frac{1}{2}\log\mu_{\cal A}-\frac{2\pi 
c}{12}t\ ,
\]
where $c\equiv (c_+ + c_-)/2$ is the average of the central charges 
$c_{\pm}$ of $\A_{\pm}$.
\end{proposition}
\proof
We have the asymptotic equality as $t\to 0^+$:
\begin{align*}
	\Tr(e^{-2\pi tH}) =& \sum_{i,j} Z_{ij} 
	\Tr(e^{-2\pi tL^+_{0,i}})\Tr(e^{-2\pi tL^-_{0,j}})\\
	\sim & \sum_{i,j} Z_{ij}d(\r^{+}_i)d(\r^{-}_j)\Tr(e^{-2\pi 
	tL^+_{0}})\Tr(e^{-2\pi tL^-_{0}})\\
	=& [\A:\A_0]\Tr(e^{-2\pi tL^+_{0}})\Tr(e^{-2\pi tL^-_{0}})\ ,
\end{align*}
where we have used the Kac-Wakimoto formula in the first equality,
while the identity $[\A:\A_0]=\sum_{i,j} Z_{ij}d(\r^{+}_i)d(\r^{-}_j)$
follows because $\bigoplus_{i,j} Z_{ij}\r^+_i\otimes \r^-_j$ is
equivalent to the canonical endomorphism of $\A_0\subset\A$, thus
\[
[\A:\A_0]=d(\sum_{i,j} Z_{ij}\r^+_i\otimes \r^-_j)
=\sum_{i,j} Z_{ij}d(\r^{+}_i)d(\r^{-}_j)\ .
\]
By \cite[Prop. 24]{KLM} we have the equality
\begin{equation}\label{2mu}
[\A:\A_0] = \sqrt{\mu_{\A_0}/\mu_{\A}} \ .
\end{equation}
Note that the above $\mu$-indices are two-dimensional, while the formula in 
\cite{KLM} concerns nets on $S^1$, but the same argument entails the 
equality \eqref{2mu}.

Therefore we have 
\[
\log\Tr(e^{-2\pi tH})\sim \frac12\big(\log\mu_{\A_0}-\log\mu_{\A}\big)
+\log\Tr(e^{-2\pi tL^+_{0}})+\log\Tr(e^{-2\pi tL^-_{0}})\ .
\]
By Prop. \ref{K} we then obtain
\begin{align*}
\log &\Tr(e^{-2\pi tH})\\
\sim & \frac12\big(\log\mu_{\A_0}-\log\mu_{\A}\big)+
\frac{\pi c_+}{12t}-\frac{1}{2}\log\mu_{\cal A_+}-\frac{\pi c_+ t}{12}
+\frac{\pi c_-}{12t}-\frac{1}{2}\log\mu_{\cal A_-}-\frac{\pi c_- t}{12}\\
\sim &\frac12\big(\log\mu_{\A_0}-\log\mu_{\A}\big)
+\frac{2\pi c}{12t}-\frac{1}{2}\log\mu_{\A_0}-\frac{2\pi c t}{12}\\
=& \frac{2\pi c}{12t}-\frac{1}{2}\log\mu_{\cal A}-\frac{2\pi c t}{12}\ ,
\end{align*}
where we have made use of the identity 
$\mu_{\A_0}=\mu_{\A_+}\mu_{\A_-}$.
\endproof
In the physical context, the expansion in Prop. \ref{2as} is natural to be 
considered, rather than the one for the chiral components in Prop. 
\ref{K}. 

Note also that a modular net $\A$ on the two-dimensional Minkowski space is 
maximal if and only if $\log\mu_\A = 0$ \cite{KL2}. This is 
consistent with the appearance of $\log\mu_\A$ only as a first order 
correction to the entropy.
\appendix
\section{Appendix. Conformal nets on $S^1$}
\label{nets}
We recall here some basic facts and results about conformal nets in 
the form needed in the paper.

We denote by $\I$ the family of proper intervals of $S^1$. 
A {\it net} $\A$ of von Neumann algebras on $S^1$ is a map 
\[
I\in\I\to\A(I)\subset B(\H)
\]
from $\I$ to von Neumann algebras on a fixed Hilbert space $\H$
that satisfies:
\begin{itemize}
\item[{\bf A.}] {\it Isotony}. If $I_{1}\subset I_{2}$ belong to $\I$, then
\begin{equation*}
 \A(I_{1})\subset\A(I_{2}).
\end{equation*}
\end{itemize}
The net $\A$ is called {\it local} if it satisfies:
\begin{itemize}
\item[{\bf B.}] {\it Locality}. If $I_{1},I_{2}\in\I$ and $I_1\cap 
I_2=\emptyset$ then 
\begin{equation*}
 [\A(I_{1}),\A(I_{2})]=\{0\},
 \end{equation*}
where the brackets denote the commutator.
\end{itemize}
The net $\A$ is called {\it M\"{o}bius covariant} if it satisfies in addition
the following properties {\bf C,D,E}:
\begin{itemize}
\item[{\bf C.}] {\it M\"{o}bius covariance}. 
There exists a strongly 
continuous unitary representation $U$ of of the M\"obius group $\Mob$ on $\H$ such that
\begin{equation*}
 U(g)\A(I) U(g)^*\ =\ \A(gI),\quad g\in\Mob,\ I\in\I.
\end{equation*}
Here $\Mob$ acts on $S^1$ by M\" obius transformations.
\item[{\bf D.}] {\it Positivity of the energy}. The generator of the 
one-parameter rotation subgroup of $U$ (conformal Hamiltonian) is positive. 
\item[{\bf E.}] {\it Existence of the vacuum}. There exists a unit 
$U$-invariant vector $\Omega\in\H$ (vacuum vector), and $\Omega$ 
is cyclic for the von Neumann algebra $\bigvee_{I\in\I}\A(I)$.
\end{itemize}
Let $\A$ be a M\"{o}bius covariant net. By the Reeh-Schlieder 
theorem the vacuum vector $\Om$ is cyclic and separating for each 
$\A(I)$. The Bisognano-Wichmann property then holds, see
\cite{BGL}: the Tomita-Takesaki modular operator $\Delta_I$ and 
conjugation $J_I$ associated with $(\A(I),\Omega)$, $I\in\I$, 
are given by 
\begin{equation}\label{BW} 
U(\Lambda_I (2\pi 
t))=\Delta_{I}^{it},\ t\in\mathbb R,\qquad U(r_I)= J_I \ . 
\end{equation} 
Here $\Lambda_I$ is the one-parameter subgroup of $\Mob$ of special 
conformal transformations preserving $I$ (also called dilatations 
associated with $I$): by identifying the upper 
semicircle $S^1$ with $\mathbb R\cup\{\infty\}$ via the stereographic 
map, thus $S^+$ with $\mathbb R^+$, $\Lambda_{S^+}(t)$ is the map $x\mapsto 
e^{-t}x$ on $\mathbb R\cup\{\infty\}$.  Then $\Lambda_{I}(t)$ is defined 
for any $I\in\I$ by conjugation by an element of $\Mob$.  $U(r_I)$ 
implements a geometric action on $\A$ corresponding to the M\" obius 
reflection $r_I$ on $S^1$ mapping $I$ onto $I^\prime$, i.e.  fixing 
the boundary points of $I$, see \cite{BGL}.  Here $I'$ denotes the 
complement of $I$, $I'\equiv S^1\setminus I$

This immediately implies Haag duality: 
\[
\A(I)'=\A(I'),\quad I\in\I\ ,
\]
where $\A(I)'$ is the commutant of $\A(I)$.

We shall say that a M\"{o}bius covariant net $\A$ is {\it irreducible} if 
$\bigvee_{I\in\I}\A(I)=B(\H)$. Indeed $\A$ is irreducible iff
$\Om$ is the unique $U$-invariant vector (up to scalar multiples), and 
iff the local von Neumann 
algebras $\A(I)$ are factors. In this case they are III$_1$-factors 
(unless $\A(I)=\mathbb C$ identically), see \cite{GL2}.

Every M\"{o}bius covariant net $\A$ decomposes uniquely into a direct 
integral of irreducible M\"{o}bius covariant nets (and the analogous is 
true for the conformal nets below); we shall thus always assume the 
following.

\begin{itemize}
\item[{\bf F.}] {\it Irreducibility}. The net $\A$ is irreducible.
\end{itemize}
Let $\Diff(S^1)$ be the group of orientation-preserving smooth 
diffeomorphisms of $S^1$. As is well known $\Diff(S^1)$ is an infinite 
dimensional Lie group whose Lie algebra is the Virasoro algebra.

By a {\it conformal net} (or diffeomorphism covariant 
net)  $\A$ we shall mean a M\"{o}bius covariant 
net such that the following holds:
\begin{itemize}
\item[{\bf G.}] {\it Conformal covariance}. There exists a projective unitary 
representation $U$ of $\Diff(S^1)$ on $\H$ extending the unitary 
representation of $\Mob$ such that for all $I\in\I$ we have
\begin{gather*}
 U(g)\A(I) U(g)^*\ =\ \A(gI),\quad  g\in\Diff(S^1), \\
 U(g)xU(g)^*\ =\ x,\quad x\in\A(I),\ g\in\Diff(I'),
\end{gather*}
where $\Diff(I)$ denotes the group of 
smooth diffeomorphisms $g$ of $S^1$ such that $g(t)=t$ for all $t\in I'$.
\end{itemize}

We shall say that $\A$ satisfies the {\it split} property if the von 
Neumann algebra
$\A(I_1)\vee\A(I_2)$ is naturally isomorphic to $\A(I_1)\otimes\A(I_2)$ 
when $I_1$ and $I_2$ are intervals with disjoint closures. The split 
property is entailed by the trace class condition 
$\Tr(e^{-tL_0})<\infty$ for all $t>0$, where $L_0$ is the conformal 
Hamiltonian.

\smallskip
\noindent
{\bf Representations}.
With $\A$ a local conformal net, 
a representation $\pi$ of $\A$ on a Hilbert space $\H$ is a map 
$I\in\I\mapsto\pi_I$ that associates to each $I$ a normal 
representation of $\A(I)$ on $B(\H)$ such that
\[
\pi_{\tilde I}\res\A(I)=\pi_I,\quad I\subset\tilde I, \quad 
I,\tilde I\subset\I\ .
\]
$\pi$ is said to be M\"obius (resp. diffeomorphism) covariant if 
there is a projective unitary representation $U_{\pi}$ of $\Mob$ (resp. 
$\Diff^{(\infty)}(S^1)$) on $\H$ such that
\[
\pi_{gI}(U(g)xU(g)^*) =U_{\pi}(g)\pi_{I}(x)U_{\pi}(g)^*
\]
for all $I\in\I$, $x\in\A(I)$ and $g\in\Mob$ (resp. 
$g\in\Diff^{(\infty)}(S^1)$). Note that if $\pi$ is irreducible and 
diffeomorphism covariant then $U$ is indeed a projective unitary 
representation of $\Diff(S^1)$.

Following \cite{DHR}, given an interval $I$ and a representation $\pi$ 
of $\A$, there is an endomorphism of $\A$ localized in $I$ equivalent 
to $\pi$; namely $\r$ is a representation of $\A$ on the vacuum Hilbert 
space $\H$, unitarily equivalent 
to $\pi$, such that $\r_{I'}=\text{id}\restriction_{\A(I')}$. We refer 
to \cite{GL2} for basic facts on this structure, in particular for 
the definition of the dimension $d(\r)$, that turns out to be equal to
the square root of the Jones index \cite{L5}.

Let $h_{\pi}$ be the conformal weight of the representation $\pi$, 
namely the lowest eigenvalue of the conformal Hamiltonian $L_{0,\pi}$ 
in the representation $\pi$. We shall need the following elementary 
fact.
\begin{lemma}\label{hl}
Let $\A$ be a local M\"{o}bius covariant conformal net on $S^1$ and 
$\pi$ an irreducible representation with $h_{\pi}=0$. 
Then $\pi$ is equivalent to the identity representation.
\end{lemma}
\proof
Let $\xi$ be a unit vector such that $L_{0,\pi}\xi = 0$. Then 
$U_{\pi}(g)\xi=\xi$ for all $g\in\Mob$ (see e.g. \cite{GL2}). Moreover, as $\pi$ is 
irreducible, $\xi$ is cyclic for $\pi$. 

Given an interval $I\in\I$ and $g_t\equiv\Lambda_I(t)$, ($t\in\mathbb 
R$), we have for every $x\in\A(I)$,
\[
(\pi_I(x)\xi,\xi)=(U_{\pi}(g_t)\pi_I(x)U_{\pi}(g_t)^{-1}\xi,\xi)=
(\pi_I(U(g_t)xU(g_t)^{-1})\xi,\xi) \ .
\]
As $t\to\infty$,  $U(g_t)xU(g_t)^{-1}$ weakly converges to 
$(x\Om,\Om)$, hence we have
\[
(\pi_I(x)\xi,\xi)=(x\Om,\Om),\quad x\in\A(I)\ ,
\]
yielding the statement by the uniqueness of the GNS representation.
\endproof

\smallskip\noindent
{\bf Nets in a non-vacuum representation.}
Given a conformal net $\A$ as above and a 
representation $\pi$ of $\A$ on a Hilbert space $\H_{\pi}$, the map
\[
I\in\I\mapsto \A_{\pi}(I)\subset B(\H_{\pi})
\]
with $\A_{\pi}(I)\equiv \pi_I(\A(I))$ satisfies all the above 
properties {\bf A} to {\bf G} (with $\A_{\pi}$ and $U_{\pi}$ in place 
of $\A$ and $U$), except {\bf E}. We can however generalize {\bf E} 
to {\bf E$'$} here below. 

A locally normal state $\o$ on $\A_{\pi}$ is, by definition,
a family $\{\o_I,\ I\in \I\}$, where $\o_I$ is a normal state on 
$\A_{\pi}(I)$, such that 
\[
\o_{\tilde I}\res\A_{\pi}(I)=\o_I  \quad {\rm if}\quad I\subset\tilde 
I\ .
\]
\begin{itemize}
\item[{\bf E$'$.}] {\it Existence of the vacuum state}. There exists a 
locally normal state $\o$ on $\A_{\pi}$ that is $\Mob$ covariant:
\begin{equation*}
 \o_I=\o_{gI}\cdot\Ad U_{\pi}(g), \quad I\in\I, \ g\in \Mob \ . 
\end{equation*}
\end{itemize}
The state $\o$ is defined by $\o_I\equiv 
(\pi_I^{-1}(\cdot)\Omega,\Omega)$ once we start with a the vacuum 
representation, but {\bf E$'$} can be taken as an axiom if we start 
directly in the representation $\pi$. In this case, in order to 
obtain the vacuum representation, one can perform the GNS procedure 
associated with $\o$. One needs however to supplement {\bf E$'$} to the 
positivity of the energy in the vacuum state, namely $\o$ must be a 
ground state. Equivalently one can require the local KMS property, 
that follows immediately from the above discussed Bisognano-Wichmann 
property if we had started from the vacuum sector.
\begin{itemize}
\item[{\bf E$''$.}] {\it Local KMS property}. The modular group 
associated with $(\A_{\pi}(I),\o_I)$, $I\in\I$, is $\Ad U_\pi(\Lambda_I(-2\pi 
t))$.
\end{itemize}
By definition a local $\Mob$ covariant net $\A_\pi$ (in an representation) 
is a map $I\in\I\mapsto\A_{\pi}(I)$ that satisfies the properties {\bf 
A,B,C,D} 
and {\bf E$'$,E$''$}. We shall say that $\A_{\pi}$ is conformal if it 
satisfies {\bf G} and the vacuum representation is diffeomorphism 
covariant.
\begin{proposition}
Let $\A_\pi$ be a local {\rm $\Mob$} covariant net in a representation. There 
exists a local {\rm $\Mob$} covariant net $\A$ in the vacuum representation 
and a DHR representation $\pi$ of $\A$ such that $\A_{\pi}(I)=\pi_I(\A(I))$.
\end{proposition}
\proof
Let $\{\H_I,\s_I,\Om_I\}$ be the GNS triple associated with $\o_I$ 
and $\A(I)\equiv\s_I(\A_\pi)$. Clearly, if $I\subset \tilde I$, we can 
identify $H_I$ with a Hilbert subspace of $\H_{\tilde I}$ and $\Om_I$ 
with $\Om_{\tilde I}$. The usual Reeh-Schlieder analyticity argument 
with the KMS property {\bf E$''$} then shows that indeed 
$\H\equiv H_I=\H_{\tilde I}$, thus $\H$ is independent of $I$. The 
rest is now clear (cf. \cite{GL5}).
\endproof
\section{Appendix. Trace and determinants}\label{App.Trace}
This appendix contains elementary known facts. Its purpose is to make 
explicit formula \eqref{logasymp2}, as it helps to understand our 
definitions.

Let $\H$ be an Hilbert space and 
$\Ga_{\pm}(\H)$ the Bose/Fermi Fock Hilbert space over $\H$.  If 
$a\in B(\H)$ and $||a||\leq 1$ the second quantization of 
$A_{\pm}\equiv\Ga_{\pm}(a)$ is the linear contraction on $\Ga_{\pm}(\H)$ defined by
\[
A_{\pm}\equiv 1\oplus a\oplus 
(a\otimes a)\oplus(a\otimes a\otimes a)\oplus\cdots
\]
where the $a\otimes\cdots\otimes a$ acts on the symmetric/anti-symmetric 
part $\H_{\pm}^{\otimes^n}$ of $\H\otimes\cdots\otimes \H$ depending on the 
Bose/Fermi alternative.
The following is well known, see e.g. \cite{BR}.
\begin{lemma}\label{9} 
If $a$ is selfadjoint, $0\leq a< 1$, then
\begin{gather}
\Tr A_{\pm}=\det(1\mp a)^{\mp 1},\\
\log\Tr A_{\pm}=\pm \Tr\log(1\pm a)
\end{gather}
\end{lemma}
\proof 
Assume first that $\H$ is one-dimensional, thus $a=\l$ is a 
scalar $0\leq\l<1$. In the Bose case, $\H_{+}^{\otimes^n}$ is also 
one-dimensional for all $n$, thus we have $A_+=\oplus_{n=0}^{\infty}\l^n$, so 
$\Tr A_+=\sum_{n=0}^{\infty}\l^n=(1-\l)^{-1}$.

For a general $a$ (with discrete spectrum) 
we may decompose $\H=\oplus_i\H_i$ so that dim$\H_i =1$ 
and $a=\oplus_i \l_i$. Then $\Ga_+(\H)=\otimes_i^{\{\Om_i\}} \Ga_+(\H_i)$, where
$\Om_i$ is the vacuum vector of $\Ga_+(\H_i)$, and  
$A_+=\otimes_i \Ga_+(a_i)$. It follows that
\[
\Tr A_+=\prod_i \Tr\Ga_+(a_i)=\prod_i (1-\l_i)^{-1} =\det(1-a)^{-1}.
\]
As for the Fermi case, if $\H$ is one-dimensional then
$\H_{-}^{\otimes^n}=\{0\}$ if $n\geq 2$ and is one-dimensional if 
$n=0,1$; if $a=\l$ we then have $A_- = 1\oplus\l$ so  $\Tr A_{-} = 1 + 
\l$. Since, also in the Fermi case, there is a 
canonical equivalence between $\Ga_-(a\oplus b)$ and 
$\Ga_-(a)\otimes\Ga_-(b)$, we have  
\[
\Tr A_{-}=\prod_i \Tr\Ga_-(\l_i)=\prod_i (1+\l_i) =\det(1+a)\ ,
\]
where $a=\oplus_i\l_i$.

Concerning the second formula, notice that 
\[
\det a=e^{\Tr\log a},
\]
hence
\[
\log\Tr A_{\pm}=\mp\log\det(1\mp a)=\mp \Tr\log(1\mp a).
\]
\endproof
\begin{lemma}\label{logasymp}
Let $h$ be a positive selfadjoint 
operator on $\H$ and $H$ the Fermi second quantization of $h$, namely 
$H=\Ga_-(h)$. Then
\begin{equation}\label{logasymp2}
\frac{\log\Tr(e^{-t H})}{\Tr(e^{-t h})}
=O(t)\qquad t\to 0^+\ .
\end{equation}
\end{lemma}
\proof 
We shall show that
\[
\log2\leq\liminf_{t\to 0^+}\frac{\log\Tr(e^{-t H})}{\Tr(e^{-t h})}
\leq\limsup_{t\to 0^+}\frac{\log\Tr (e^{-t H})}{\Tr(e^{-t h})}=1.
\]
By Lemma \ref{9} it suffices to show that
\[
\log2\leq\liminf_{t\to 0^+}\frac{\Tr\log(1+e^{-t h})}{\Tr(e^{-t h})}
\leq\limsup_{t\to 0^+}\frac{\Tr\log(1+e^{-t h})}{\Tr(e^{-t h})}=1.
\]
We have
\[
\log 2\cdot e^{-t h}\leq \log(1+e^{-t h}) \leq e^{-t h}
\]
because of the corresponding function inequalities, that obviously implies 
the previous inequality.
\endproof
The Bose version of the above lemma is omitted (the $U(1)$-current 
algebra local conformal net is not rational).
\section{Appendix. Index and entropy}
\label{math}
In this appendix we develop abstract mathematical results, 
concerning Jones index and Connes-Haagerup noncommutative measure 
theory, that are necessary for our work. We refer to Takesaki's book 
\cite{T} for the basic theory.
 
Let $ R $ be a von Neumann algebra on a Hilbert space $\H$, $S=R'$ its 
commutant.  Given a n.f.s.  (normal, faithful, semifinite) weight $\f$ 
on $R$ and a n.f.s.  weight $\psi$ on $S$ the Connes spatial 
derivative ${\frac{\text{d} \f}{\text{d}\psi}}$ is a canonical 
positive non-singular selfadjoint operator on $\H$ such that 
$\big({\frac{\text{d} \f}{\text{d}\psi}}\big)^{it}$ implements 
$\s_t^{\f}$ on $ R$ (the modular group of $(R,\f)$) and 
$\big({\frac{\text{d} \f}{\text{d}\psi}}\big)^{-it}$ implements 
$\s_t^{\psi}$ on $S$.  One has ${\frac{\text{d} \f}{\text{d}\psi}}= 
\big({\frac{\text{d} \psi}{\text{d}\phi}}\big)^{-1}$.

If $\psi_0$ is another n.f.s. weight on $S$ there holds
\begin{equation}
\label{chain0}
\big({\frac{\text{d}\f}{\text{d} \psi_0}}\big)^{it}=
\big({\frac{\text{d}\f}{\text{d} 
\psi}}\big)^{it}(D\psi : D\psi_0)_t  \ ,
\end{equation}
where $(D\psi : D\psi_0)$ is the unitary Connes Radon-Nikodym cocycle in $S$ w.r.t. 
$\psi_0$ and $\psi$.
The following proposition is known.
\begin{proposition}\label{CH}
Let $ R$ and $S=R'$ be von Neumann algebras on a Hilbert space $\H$,
and and $V$ a one-parameter 
unitary group on $\H$ such that $\Ad V(t)R=R$, $t\in\mathbb R$.
Given a n.f.s. weight $\f$ on $R$ such 
that $\Ad V(t)\restriction_{R} = \s^\f_t$, there is 
a unique n.f.s. $\psi$ weight on $S$ such that 
$\big(\frac{{\rm d} \f}{{\rm d}\psi}\big)^{it}=V(t)$.

If $\psi_0$ is an arbitrary n.f.s. weight on $S$ one has
\begin{equation}
\label{chain}
(D\psi : D\psi_0)_t = u_t
\end{equation}
where $u_t \equiv 
V(-t)\big({\frac{{\rm d}\f}{{\rm d}\psi_0}}\big)^{it}$.
\end{proposition}
\proof
With $\psi_0$ an arbitrary n.f.s. weight on $S$, both 
$\big({\frac{\text{d} \f}{\text{d}\psi_0}}\big)^{it}$ and $V(t)$ 
implements $\s^{\f}_t$ on $R$, thus $u_t$ belongs to $S$ and is a 
unitary $\s^{\psi_0}$-cocycle. By Connes theorem, there 
exists a n.f.s. weight $\psi$ on $S$ such that $u_t = (D\psi : 
D\psi_0)_t$. The rest follows by formula (\ref{chain0}).
\endproof
\begin{corollary}\label{xi}
Suppose that, in Prop. \ref{CH}, $\f$ is the state on $R$ given by 
a cyclic and separating vector $\xi$. If $K$ is the infinitesimal 
generator of $V$ we have
\[
\psi(1)=(e^{-K}\xi,\xi),
\]
in particular $\psi$ is a bounded functional iff $\xi$ belongs to the 
domain of $e^{-{\frac12}K}$.
\end{corollary}
\proof
Let $\psi_0$ be the vector state on $S$ implemented by $\xi$. Then
\begin{multline}
\psi(1)=\aci \psi_0\big((D\psi : D\psi_0)_t\big)\\=
\aci (V(-t)\big({\frac{\text{d}\f}{\text{d}\psi_0}}\big)^{it}\xi,\xi)=
\aci (V(-t)\xi,\xi)= (e^{-K}\xi,\xi)
\end{multline}
where we have made use that ${\frac{\text{d}\f}{\text{d}\psi_0}}$ is 
the modular operator of $(R,\xi)$, thus 
${\frac{\text{d}\f}{\text{d}\psi_0}}\xi=\xi$.
\endproof
Let $N_1$, $N_2$ be commuting factors on a Hilbert space $\H$ with 
$N_1\vee N_2=B(\H)$. Set $M_1\equiv N'_2$, $M_2\equiv N'_1$, thus 
$N_i \subset M_i$ are irreducible inclusion of factors ($i=1,2$).
Let $\f_i$ be a normal faithful state on $N_i$ and $V$ a one-parameter 
unitary group on $\H$ such that
\[
\Ad V(t)\restriction_{N_1} = \s^{\f_1}_t, \quad \Ad 
V(-t)\restriction_{N_2} = 
\s^{\f_2}_t, \quad t\in \mathbb R,
\]
where $\s^{\f_i}$ is the modular group of $(N_i,\f_i)$.
Let $\psi_1$ be the n.f.s. weight on $M_1$ associated 
with $V$ and $\f_2$ by Prop. \ref{CH}, namely $\psi_1$ is 
characterized by
\[
K = \log\big(\frac{\text{d} \f_2}{\text{d} \psi_1}\big)\ ,
\]
and analogously let $\psi_2$ be the n.f.s. weight on $M_2$ associated
with $V$ and $\f_1$.

There exists a unique n.f.s.  operator valued weight $\E_i: M_i\to 
N_i$ such that $\f_i\cdot\E_i =\psi_i$.  The existence of $\E_i$ follows 
by Haagerup theorem because $\s^{\psi_i}\restriction_{N_i} = \s^{\f_i}$.  
Then $\E_i$ is faithful and unique up to a positive scalar multiple 
because $N'_i\cap M_i=\mathbb C$.
\begin{proposition}\label{OVW}
The following are equivalent:
\begin{itemize}
\item[$(a)$] There exists a normal expectation $\e_i:M_i\to N_i$; 
\item[$(b)$] $\psi_i$ is bounded.
\end{itemize}
If the above hold, then $\E_i=\psi_i(1)\e_i$ and 
\begin{align}
K =& -\log\frac{{\rm d} \f_1\cdot\e_1}{{\rm d} \f_2} + \log\psi_1(1)
\\
=& \log\frac{{\rm d} \f_2\cdot\e_2}{{\rm d} \f_1} + \log\psi_2(1)
\ .\label{Kdel}
\end{align}
\end{proposition}
\proof
If $(a)$ holds, say with $i=1$, then $\E_1=\lambda\e_1$ for some $\l>0$, thus 
$\psi_1=\f_1\cdot\E_1= \psi_1=\l\f_1\cdot\e_1$ is bounded. Conversely if 
$(b)$ holds then $\psi_1$ is a normal, faithful, positive linear 
functional on $M_1$ whose modular group $\s_t^{\psi_1}=\Ad V(t)$ 
leaves $N_1$ globally invariant, so there is a normal expectation 
$\e:M_1\to N_1$ by Takesaki theorem.

Clearly, if the above hold, then $\E_1(1)=\l$, thus
$\psi_1(1)=\f_1\cdot\E_1(1)=\l$, and the rest of the
statement follows.
\endproof
Assume there exists a faithful normal expectation $\e_1:M_1\to N_1$. 
Denote by $\e^{-1}:M_2\to N_2$ the dual operator valued weight. This is 
the unique n.f.s. operator valued weight $M_2\to N_2$ such that
\[
\frac{\text{d} \o_1\cdot \e_1}{\text{d} \o_2}= 
\big(\frac{\text{d} \o_2\cdot \e^{-1}}{\text{d} \o_1}\big)^{-1}
\]
for all n.f.s. weight $\o_1$ on $N_1$ and $\o_2$ on $N_2$.

According to Kosaki definition \cite{Ko}, the inclusion $N_1\subset M_1$ has 
finite index iff $\e^{-1}$ is bounded and the index is defined to be 
$\e^{-1}(1)$, namely
\[
\e^{-1}=[M_1 : N_1]\e_2\ ,
\] 
where $\e_2$ is the unique normal expectation from $M_2$ onto $N_2$.
\begin{proposition}\label{psi}
We have
\[
[M_1:N_1] = \psi_1(1)\cdot\psi_2(1)\ .
\]
\end{proposition}
\proof
By definition
\[
\frac{\text{d} \psi_1}{\text{d} \f_2} = e^{K},\quad \frac{\text{d} \psi_2}{\text{d} \f_1} = 
e^{-K} \ .
\]
Thus 
\[
\frac{\text{d} \f_1\cdot\E_1}{\text{d} \f_2}= \big(\frac{\text{d} \f_2\cdot\E_2}{\text{d} 
\f_1}\big)^{-1}\ ;
\]
setting $\l_i\equiv\psi_i(1)$, since $\E_i=\l_i\e_i$ we then have
\[
\l_1\l_2\frac{\text{d} \f_1\cdot\e_1}{\text{d} \f_2}= 
\big(\frac{\text{d} \f_2\cdot\e_2}{\text{d} 
\f_1}\big)^{-1}\ .
\]
On the other hand we have
\[
\frac{\text{d} \f_1\cdot\e_1}{\text{d} \f_2}= 
[M_1:N_1]^{-1}\big(\frac{\text{d}\f_2\cdot\e_2}{\text{d}\f_1}\big)^{-1}
\]
showing that $[M_1:N_1]=\l_1\l_2$.
\endproof
\begin{corollary}
\label{absf}
If $\xi_i$ is a cyclic and separating vector for $N_i$ such that 
$\f_i(x)=(x\xi_i, \xi_i)$, $x\in N_i$, we have
\[
[M_1:N_1]=(e^K \xi_1,\xi_1)(e^{-K} \xi_2,\xi_2)\ .
\]
Suppose further that there exists a unitary $U$ such that 
$UM_1 U^*=M_2$, $UN_1 U^*=N_2$, $\f_2 = \f_1\cdot\Ad U$ and 
$UV(t)U^* = V(-t)$. Then
$\psi_1(1)=\psi_2(1)=[M_1 : N_1]^{\frac12}$ and
\[
(e^K \xi_1,\xi_1)=(e^{-K} \xi_2,\xi_2)=[M_1:N_1]^{\frac12}\ ,
\]
thus
\begin{equation}\label{last}
K = -\log\frac{{\rm d}\f_1\cdot\e_1}{{\rm d}\f_2} + \frac12\log 
[M_1:N_1]\ .
\end{equation}
\end{corollary}
\proof
The first equality follows by Cor. \ref{xi} and Prop. \ref{psi}.

The second equality then follows because $U$ interchanges the triples 
of $(M_1, N_1,\f_1)$ and  $(M_2, N_2,\f_2)$, thus the canonical 
quantities $(e^K \xi_1,\xi_1)$ and $(e^{-K} \xi_2,\xi_2)$ must 
coincide.

The last identity \eqref{last} now follows by equation \eqref{Kdel}.
\endproof

\smallskip\noindent
{\bf Araki relative entropy.} Before concluding this appendix we 
recall the definition of Araki relative entropy between two faithful normal 
states $\f_1$ and $\f_2$ of  von Neumann algebra $M$:
\[
S(\f_1 | \f_2) \equiv -(\log\Delta_{\xi_2,\xi_1}\xi_1 ,\xi_1) \ .
\]
Here $M$ is in a standard form with respect to a cyclic and separating 
vector $\Om$, the vector $\xi_i$ is the canonical 
representative of $\f_i$ in the 
natural positive cone $L^2(M,\Om)_+$ and $\Delta_{\xi_2 , \xi_1}$ is 
the relative modular operator, namely the polar decomposition of 
$S_{\xi_2 , \xi_1}$ is
\[
S_{\xi_2 , \xi_1}= J\Delta_{\xi_2 , \xi_1}^{1/2}
\]
where $S_{\xi_2 , \xi_1}$ is the closure of the anti-linear operator on 
$M\xi_1$ defined by $S_{\xi_2 , \xi_1}x\xi_1 = x^*\xi_2$. 

It easy to check that $S_{\xi_2 , \xi_1} = S_{\eta_2 , \eta_1}$ if 
$\eta_1$ implements the same state of $\xi_1$ on $M$ and 
$\eta_2$ implements the same state of $\xi_2$ on $M'$ , namely
$\f_1 = (\cdot\eta_1 , \eta_1)\res_M$ and $\psi_2\equiv\f_2\cdot\Ad J = 
(\cdot\eta_2 , \eta_2)\res_{M'}$.
Thus $S(\f_1 |\f_2)$ depends only on the states $\f_1$ and $\psi_2$ 
and we have
\[
S(\f_1 | \f_2)= S(\f_1 | \psi_2)\equiv 
-(\log\!\big(\frac{\text{d}\f_1}{\text{d}\psi_2}\big)\xi_1 ,\xi_1) \ .
\]
We finally note, that, by taking expectation values, equation
\eqref{last} gives
\begin{align*}
(K\xi_2 , \xi_2) &=
-(\log\frac{{\rm d}\f_1\cdot\e_1}{{\rm d}\f_2}\xi_2 , \xi_2) + 
\frac12\log [M_1:N_1]\\
=& S(\f_2 | \f_1\cdot\e_1) + \tfrac12 H(M_1 | N_1) \ ,
\end{align*}
where $H(M_1 | N_1) = \log[M_1 : N_1]$ is the Pimsner-Popa entropy 
\cite{PP}.

\section{Final comments}
Adding a massive charge to a black hole should increase the total mass
of the black hole, hence make a change of the spacetime itself and of
the entropy.  In a theory of quantum gravity, the spacetime itself
should be noncommutative \cite{DFR} from the start.  In the setting of
QFT on a curved spacetime the backreaction from the gravitational
field is ignored and the spacetime is classical.  In the previous 
work \cite{L2} one considered the addition of a single charge: the
increment of entropy is there an ``higher order effect'' and become visible
in the associated noncommutative geometry, while the classical
spacetime remains fixed \cite{L4}.  The entropy in the present work
also has a noncommutative geometrical nature, but rather reflects the
global noncommutative geometrical complexity of the system.

It would be interesting to relate our setting with Connes'
Noncommutative Geometry \cite{Co}.  A link should be possible in a
supersymmetric context, where cyclic cohomology appears.  In this
respect model analysis with our point of view, in particular in the
supersymmetric frame, may be of interest.  Note also that Connes' 
spectral action concerns the Hamiltonian spectral density 
behavior, see \cite{CC}.
\bigskip

\noindent {\bf Acknowledgements.} The second named author wishes to
thank, among others, I.M. Singer for an initial stimulating comment
and A. Connes for a wide perspective e-mail exchange on the subject.
Thanks also to B. Schroer for comments on the final manuscript.
\end{document}